\documentclass[%
nofootinbib,
 amsmath,amssymb,
 aps,
]{revtex4-2}

\usepackage{graphicx}
\usepackage{dcolumn}
\usepackage{bm}
\usepackage{hyperref}

\begin{document}

\title{Temperature Mapping on a Niobium-Coated Copper Superconducting Radio-Frequency Cavity}

\author{Antonio Bianchi} \email{Corresponding author: antonio.bianchi@cern.ch}
\author{Walter Venturini Delsolaro}%
\affiliation{CERN}

\begin{abstract}

Since the late '80s, CERN has pioneered the development of niobium thin film radio-frequency (RF) cavities deposited on copper substrates for several particle accelerator applications. However, niobium thin film cavities historically feature a progressive performance degradation as the accelerating field increases. In this study, we describe a temperature mapping system based on contact thermometry, specially designed to obtain temperature maps of niobium-coated copper cavities and, consequently, study the mechanisms responsible for performance degradation. The first temperature maps on a niobium/copper 1.3 GHz cavity are reported along with its RF performance. In addition to some hotspots displayed in the temperature maps, we surprisingly observed a wide cold area on the outer cavity surface that may shed new light on understanding the heat dissipation of niobium thin film cavities in liquid helium, which might be exploited to improve the RF cavity performance.

\end{abstract}

\maketitle


\section{\label{sec:level1}INTRODUCTION}

Since the late ‘80s, CERN has pioneered the development of thin film radio-frequency (RF) cavities for particle accelerators. Niobium (Nb) thin film on copper (Cu) cavities have been successfully applied at CERN in the Large Electron Positron collider (LEP-II) \cite{benvenuti1991superconducting}, in the Large Hadron Collider (LHC) \cite{Bruning_782076} and more recently in the High Intensity and Energy Isotope Separator On Line DEvice (HIE-ISOLDE) \cite{miyazaki2019two}. Many efforts are now put in place at CERN in view of their potential implementation in the Future Circular Collider (FCC) machines.

Temperature mapping systems are one of the most valuable diagnostics to investigate mechanisms responsible for performance degradation in SRF cavities. The effectiveness of temperature mapping systems is to detect any loss mechanisms inside the cavity that ultimately produces heat in the cavity substrate \cite{ciovati2005temperature}. This is also useful to ensure quality control during series production of many SRF cavities. In the last decades, temperature mapping systems significantly contributed to the improvement of cavity performance  \cite{padamsee2020history, ciovati2005temperature}. Indeed, by temperature sensing on the outer surface of SRF cavities, several types of losses that occur inside the cavities can be detected, such as multipacting processes, field emission, ohmic loss mechanisms, and quenches.

The first system for temperature mapping on bulk Nb radio-frequency cavities was built at Cornell University by H. Padamsee \cite{padamsee2020history, ciovati2005temperature}. This system has been duplicated in other laboratories \cite{padamsee2020history, ciovati2005temperature, ciovati20082, pekeler1996thermometric, yamamoto2009new, koszegi2018combined}. All these setups are applied in testing bulk Nb cavities generally operated in superfluid helium (He-II), where the heat exchange from cavities to the helium (He) bath is dominated by the Kapitza resistance. In the past, only one system was developed for Nb thin film cavities deposited on Cu substrates. The system, built at CERN in the '80s, used a rotating arm of thermometers for mapping Nb/Cu 500 MHz cavities in liquid He above the lambda-point (He-I) in the subcooled condition. However, Nb/Cu cavities are usually operated in He-I at saturation pressure, where the heat transfer from cavities to He bath is different from that in He-II at saturation pressure or in He-I in the subcooled condition. Another drawback of the system was the long acquisition time, from half an hour to one hour for a complete temperature map. Consequently, only steady-state RF losses in the cavity were detected \cite{padamsee2020history}. In addition, the temperature of the subcooled He bath is not stable over time; in fact, it slowly drifts to higher values. This has to be carefully considered when the acquisition time of a temperature map is quite long.

No temperature mapping systems for Nb/Cu cavities are currently in operation, as far as we know. This paper describes a system based on contact thermometry, specifically designed for testing Nb/Cu cavities in He-I at saturation pressure and in the subcooled condition. As a result, we can report the first temperature maps of a Nb/Cu 1.3 GHz cavity, along with its RF performance. Temperature mapping of the cavity during its operation showed some interesting sites with characteristics consistent with Joule heating and field emission heating. An optical inspection of the internal cavity surface confirmed the presence of defects on the Nb thin film in correspondence with these sites. Contrary to expectations, we also observed a wide cold area on the outer surface that may shed new light on the heat dissipation of Nb/Cu cavities in He-I, and might be exploited to improve the RF cavity performance.

The paper is organized as follows. Section \ref{sec:level2} gives a brief overview of parameters that play a role in mapping heat losses on Cu surfaces at liquid He temperatures. In section \ref{sec:level3}, we describe the temperature mapping system developed for testing Nb/Cu 1.3 GHz cavities. After reporting the RF performance of the cavity under test in section \ref{sec:level4}, we examine the temperature maps acquired at 2.4 K when the He bath is at saturation pressure and in the subcooled condition in sections \ref{sec:level5} and \ref{sec:level6}, respectively. The optical inspection results of the cavity are outlined in section \ref{sec:level7}. Finally, we investigate how to improve the heat dissipation of Nb/Cu cavities into the He bath in section \ref{sec:level8}.

\section{\label{sec:level2}TEMPERATURE MAPPING ON COPPER SURFACES AT LIQUID HELIUM TEMPERATURES}

Mapping heat losses on Nb thin film cavities deposited on Cu substrates presents a greater challenge compared to bulk Nb cavities.

\begin{figure}[!htb]
   \centering
   \includegraphics*[width=0.55\columnwidth]{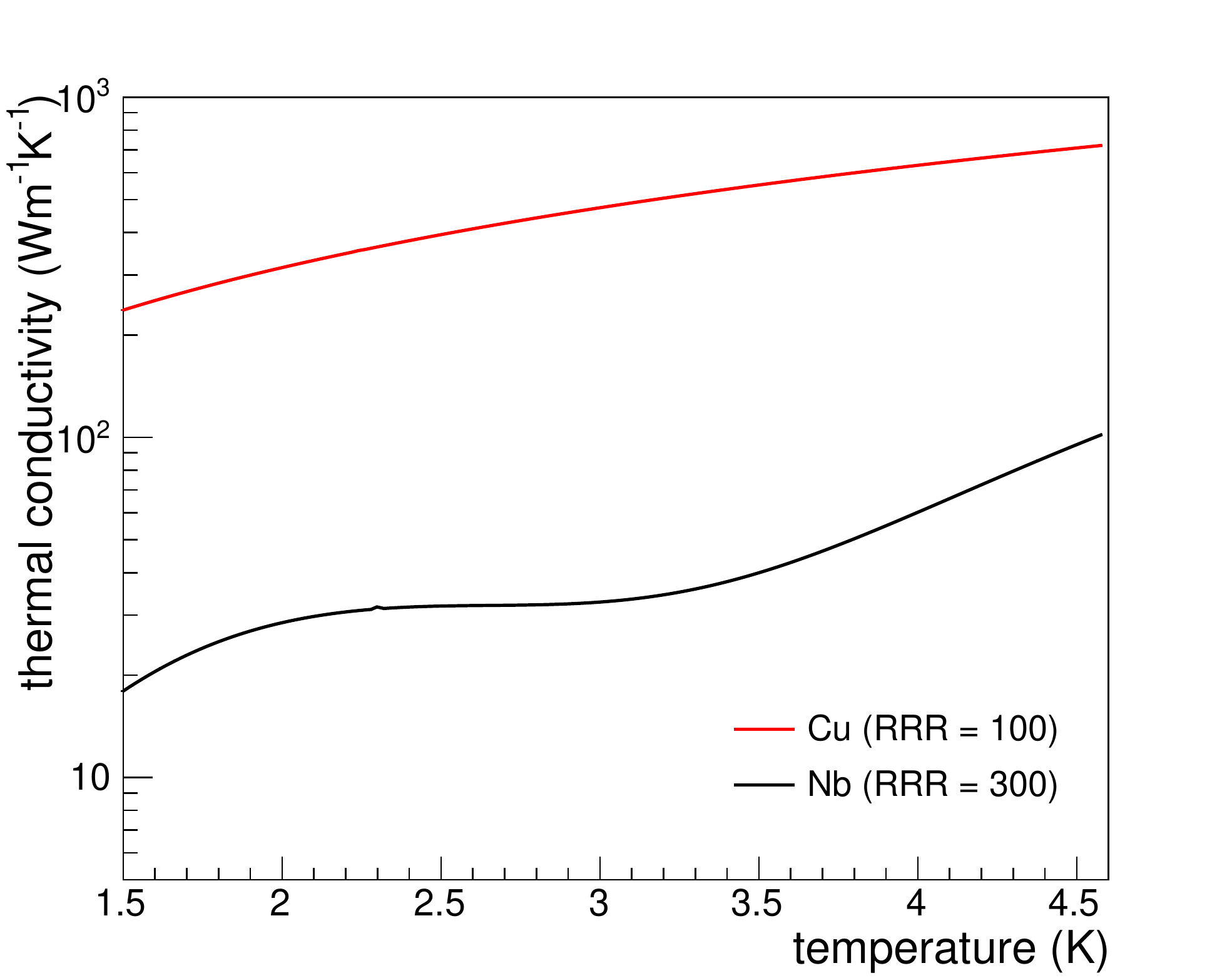} 
   \caption{Thermal conductivity of Cu (RRR = 100) \cite{russenschuck2011field} and Nb (RRR = 100) \cite{padamsee1983calculations} between 1.5 K and 4.6 K.}
   \label{fig:plot1}
\end{figure}

The thermal conductivity of Cu is more than one order of magnitude higher than that of Nb at liquid He temperatures \cite{padamsee1983calculations, russenschuck2011field}. Figure \ref{fig:plot1} shows the thermal conductivity of Cu and Nb between 1.5 K and 4.6 K, which is the temperature range where SRF cavities are usually tested. The Cu substrates, generally used for Nb/Cu cavities, have a residual resistance ratio (RRR) between 50 and 100. In contrast, the RRR value of Nb sheets used for manufacturing bulk Nb cavities is usually equal to $\sim$300 \cite{padamsee}.

\begin{figure}[!htb]
   \centering
   \includegraphics*[width=0.55\columnwidth]{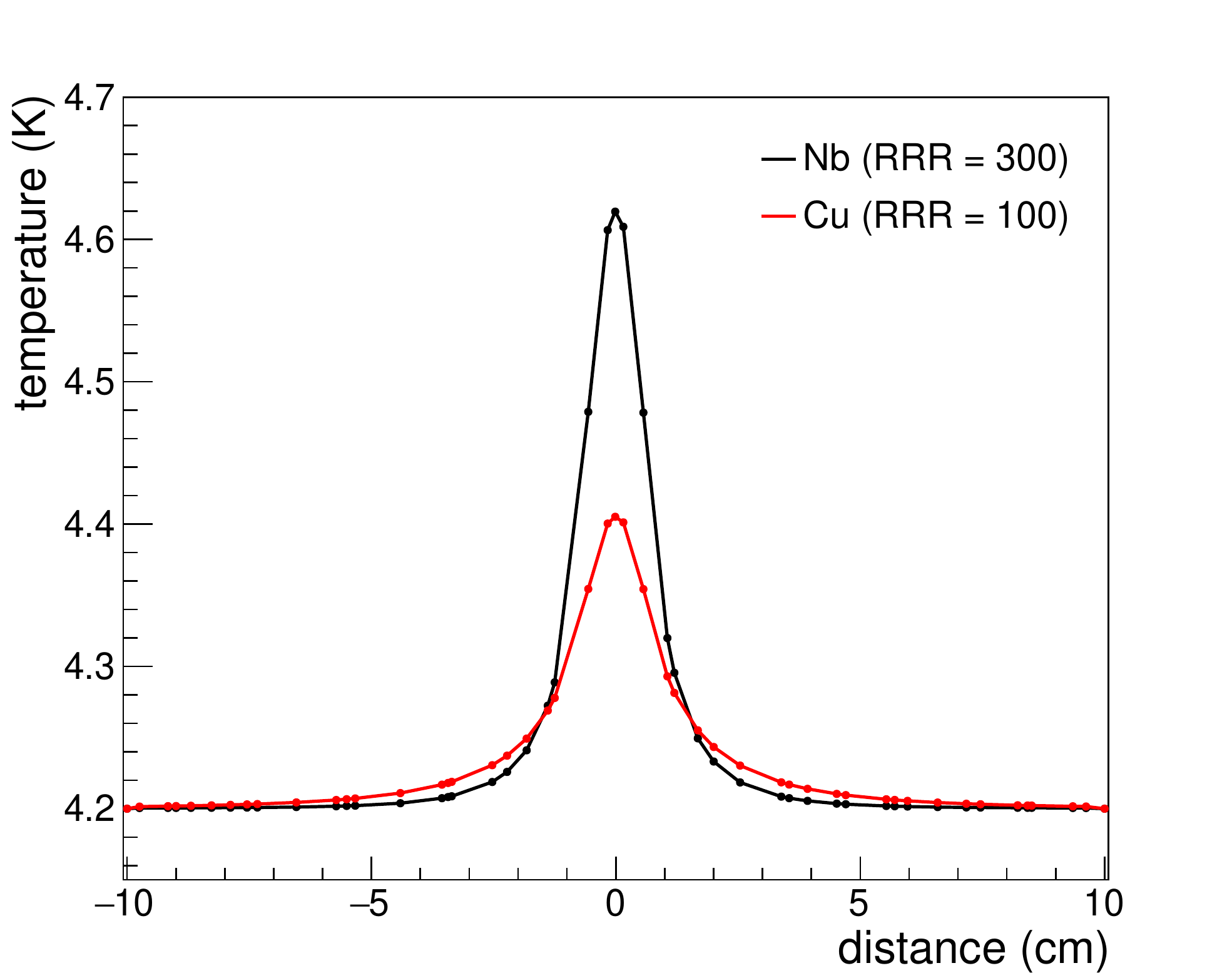} 
   \caption{Temperature profile in a disk with a diameter of 20 cm and thickness of 2 mm. The disk is heated in its center with a heat flux density of 1 W/cm$^{2}$ and cooled by liquid He at 4.2 K at saturation pressure. The temperature profile simulated in a disk in Nb (RRR = 300) is higher by a factor of 2 and slightly narrower than that simulated in a disk in Cu (RRR = 100).}
   \label{fig:plot2}
\end{figure}

The thermal conductivity of Cu is $\sim$300 W/(m$\cdot$K) at 1.8 K and $\sim$700 W/(m$\cdot$K) at 4.2 K, whereas these values in Nb are only $\sim$25 W/(m$\cdot$K) and $\sim$60 W/(m$\cdot$K), respectively \cite{padamsee1983calculations, russenschuck2011field}. This implies that the temperature profile on Cu surfaces in correspondence with a given heat loss is lower than that on bulk Nb surfaces. Figure \ref{fig:plot2} shows the temperature profile in a disk when a heat loss of 1 W/cm$^{2}$ is located at the center. The disk is cooled by liquid He at 4.2 K at saturation pressure. The disk’s diameter is 20 cm and its thickness is equal to 2 mm, which is quite similar to most of the substrates for SRF cavities. In the case of Cu, the temperature profile in the disk is lower than that of Nb by a factor of $\sim$2 and slightly broader. This makes measuring temperature profiles on Cu surfaces more complex and challenging than on bulk Nb surfaces.

Unlike bulk Nb cavities, Nb/Cu cavities are usually operated in He-I at saturation pressure. In this condition, the heat transfer from the Cu surface to the He bath occurs by convection cooling for low heat fluxes per unit surface area, whereas gaseous He bubbles appear on the Cu surfaces at higher heat fluxes, leading to the transition from convection cooling to nucleate boiling. Cooling by nucleate boiling is more effective than convection cooling in terms of heat transfer from the Cu surface to the He bath \cite{smith1969review}. The activation of nucleation sites, where bubbles of gaseous He grow, frequently implies a sudden temperature drop in the surface \cite{smith1969review}. According to results of previous studies \cite{bianchi_srf2021-thpcav007, bianchi_tobepublished}, the ideal condition for mapping heat losses on Nb/Cu cavities in He-I is slightly above the lambda-point of He, which is at $\sim$2.17 K. We verified that the He bath temperature of 2.4 K allows us to carry out measurements with satisfactory accuracy. At this temperature, the thermal conductivity of Cu is lower than that at 4.2 K, as shown in figure \ref{fig:plot1}. Furthermore, the thermal conductivity of He-I also decreases by $\sim$30\% from 4.2 K to 2.4 K \cite{osti_7031380}. As a consequence, in the presence of a point-like heat loss, the temperature profiles on Cu surfaces at 2.4 K are higher and broader than those at 4.2 K. 

A further parameter to be considered for mapping heat losses on Cu surfaces in He-I is the sensitivity of thermometers. Previous studies show that Allen-Bradley 100 $\Omega$ carbon resistors can be used as thermometers for measuring temperature profiles on Cu surfaces immersed in He-I \cite{bianchi_srf2021-thpcav007, bianchi_tobepublished}. The sensitivity of Allen-Bradley 100 $\Omega$ resistors at 2.4 K is higher than that at 4.2 K, as will be discussed in section \ref{sec:level3e1}.

Another thermodynamic state where heat losses in Nb/Cu cavities can be satisfactorily detected is the subcooled He condition \cite{bianchi_tobepublished}. Indeed, H. Piel in the ’80s observed that the temperature profiles on Cu surfaces in subcooled He-I were much higher and broader than those at saturation pressure \cite{piel1980diagnostic, piel1981cern}. This is mainly due to the reduced cooling capability of the He bath in subcooled conditions, which is lower than that at saturation pressure \cite{piel1980diagnostic, piel1981cern}. However, Nb/Cu cavities are not generally operated in subcooled He. Therefore, temperature maps of Nb/Cu cavities in He-I at saturation pressure can be used to understand their heat dissipation at operating conditions, whereas temperature mapping in subcooled He allows the localization and characterization of heat losses with high precision and accuracy.

\section{\label{sec:level3}TEMPERATURE MAPPING SYSTEM FOR NIOBIUM/COPPER CAVITIES}

A temperature mapping system has been specially developed at CERN to test Nb/Cu cavities. Based on contact thermometry, the system is used to map the temperature on the outer surface of 1.3 GHz single-cell TESLA-type cavities \cite{aune2000superconducting}. In the system, temperature sensing is by 192 thermometers made of carbon resistors. The data acquisition software is developed to provide real-time temperature maps. Therefore, the evolution of heat losses can be monitored and tracked during the cavity cold tests. The design is partially based on that developed at Jefferson Lab for bulk Nb 1.5 GHz CEBAF cavities \cite{ciovati2005temperature}. 

The complete description of our system is organized as follows. Section \ref{sec:level3e1} describes the thermometers and their cabling. Electronics is examined in section \ref{sec:level3e2}. Finally, the data acquisition software is presented in section \ref{sec:level3e3}.

\subsection{\label{sec:level3e1}Thermometers}

\begin{figure}[!htb]
   \centering
   \includegraphics*[width=0.55\columnwidth]{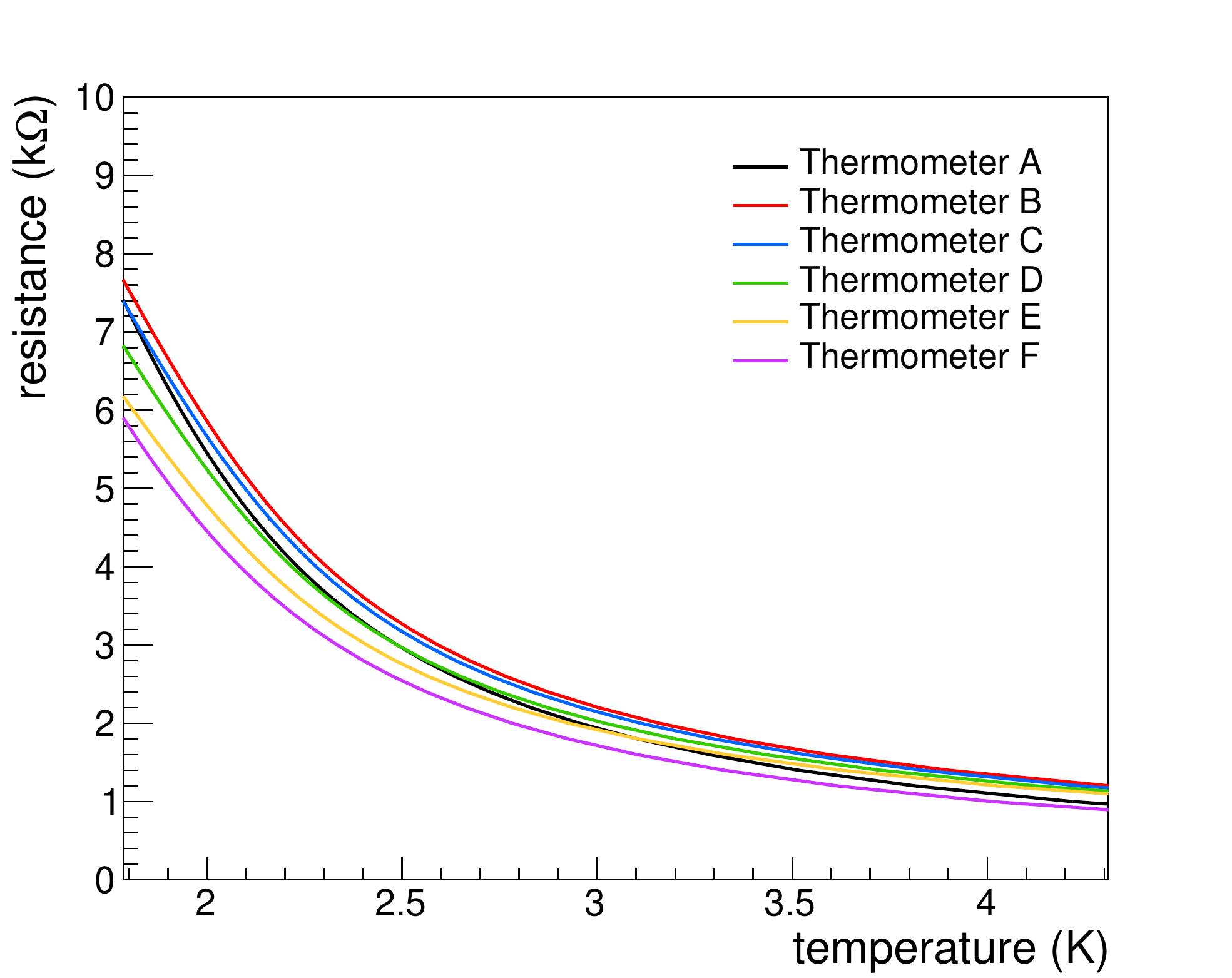}
   \caption{Resistance variation of six Allen-Bradley 100 $\Omega$ carbon resistors between 1.8 K and 4.2 K.}
   \label{fig:plot3}
\end{figure}

For the temperature mapping system, we used 192 Allen-Bradley 100 $\Omega$ carbon resistors as thermometers. The resistance of these carbon resistors increases by decreasing the temperature. Indeed, their resistance is generally $\sim$7 k$\Omega$ at 1.8 K, $\sim$3 k$\Omega$ at 2.4 K and $\sim$1 k$\Omega$ at 4.2 K. Figure \ref{fig:plot3} shows the resistance variation of six thermometers between 1.8 K and 4.2 K.

\begin{figure}[!htb]
   \centering
   \includegraphics*[width=0.55\columnwidth]{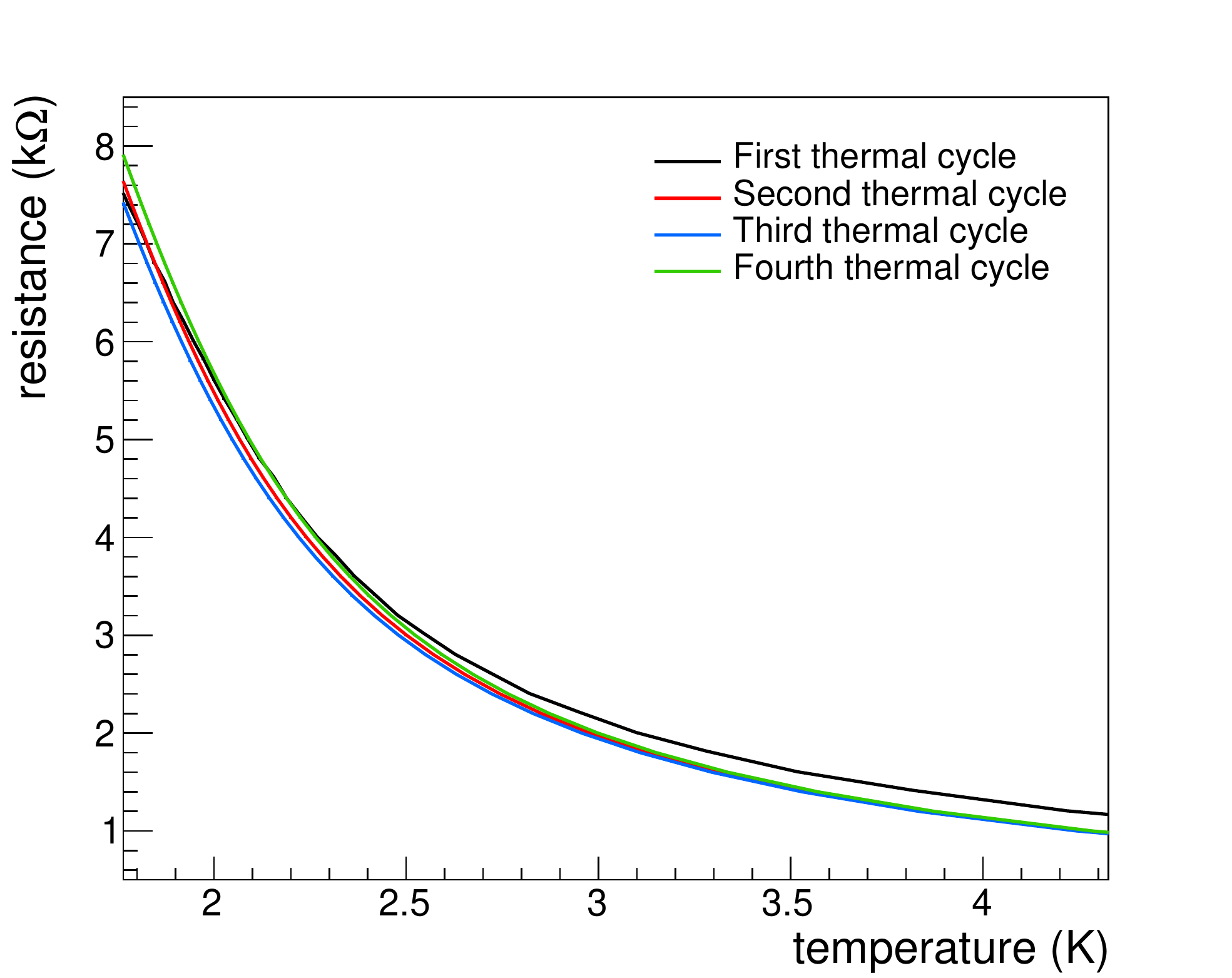} 
   \caption{Resistance variation of one Allen-Bradley 100 $\Omega$ carbon resistor in the temperature range between 1.8 K and 4.2 K for four consecutive thermal cycles at room temperature.}
   \label{fig:plot4}
\end{figure}

The resistance variation of Allen-Bradley 100 $\Omega$ carbon resistors is high enough to detect small temperature variations on Cu surfaces in liquid He-I and He-II, as already demonstrated in previous studies \cite{bianchi_srf2021-thpcav007, bianchi_tobepublished}. However, their resistance slightly changes when a thermal cycle at room temperature is performed. This implies that thermometers must always be calibrated for each system's cool-down. For example, figure \ref{fig:plot4} shows the resistance variation of one thermometer, randomly chosen, in the temperature range between 1.8 K and 4.2 K for four consecutive thermal cycles at room temperature. If no calibration procedure is carried out after each thermal cycle, the uncertainty on temperature can be as high as 150 mK \cite{bianchi_srf2021-thpcav007}.

\begin{figure}[!htb]
   \centering
   \includegraphics*[width=0.5\columnwidth]{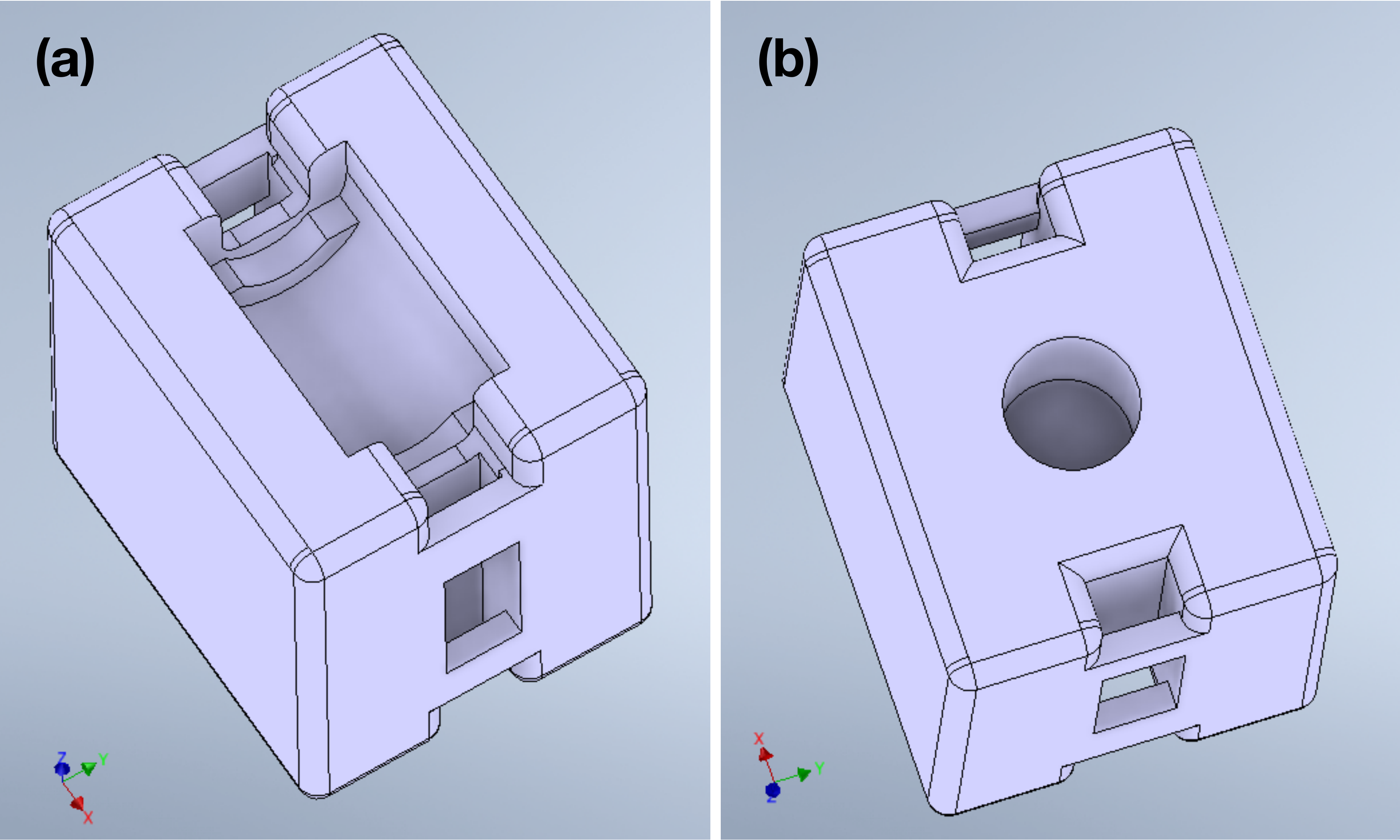} 
   \caption{Top (a) and bottom view (b) of the Accura 25 housing where each thermometer is embedded and sealed by Stycast 2850 FT.}
   \label{fig:plot5}
\end{figure}

Thermometers are embedded in an Accura 25 housing as shown in figure \ref{fig:plot5}. We verified that Accura 25 3D-printed plastic is suitable for cryogenic temperatures. The housing is sealed by Stycast 2850 FT, which has high thermal conductivity and is impervious to superfluid He. The Accura 25 housing ensures thermal isolation from the He bath, and the Stycast epoxy fully protects them from liquid He. Apiezon N grease is applied between the thermometers and the external cavity surface to lower the thermal contact resistance and, consequently, optimize the temperature measurements \cite{bianchi_srf2021-thpcav007, bianchi_tobepublished}. Thanks to the special design of the thermometer housing in Accura 25 and the use of Apiezon N grease as thermal paste, the average efficiency\footnote{The efficiency of thermometers is defined as the ratio of the measured temperature rise to the actual temperature rise of the Cu surface \cite{bianchi_tobepublished}.} of thermometers is $\sim$40\% at 2.4 K at saturation pressure for values of heat flux density ranging from 0.4 W/cm$^{2}$ to 2.3 W/cm$^{2}$. In contrast, the efficiency is $\sim$70\% at the same temperature of the He bath but in the subcooled condition \cite{bianchi_tobepublished}.

\begin{figure}[!htb]
   \centering
   \includegraphics*[width=0.4\columnwidth]{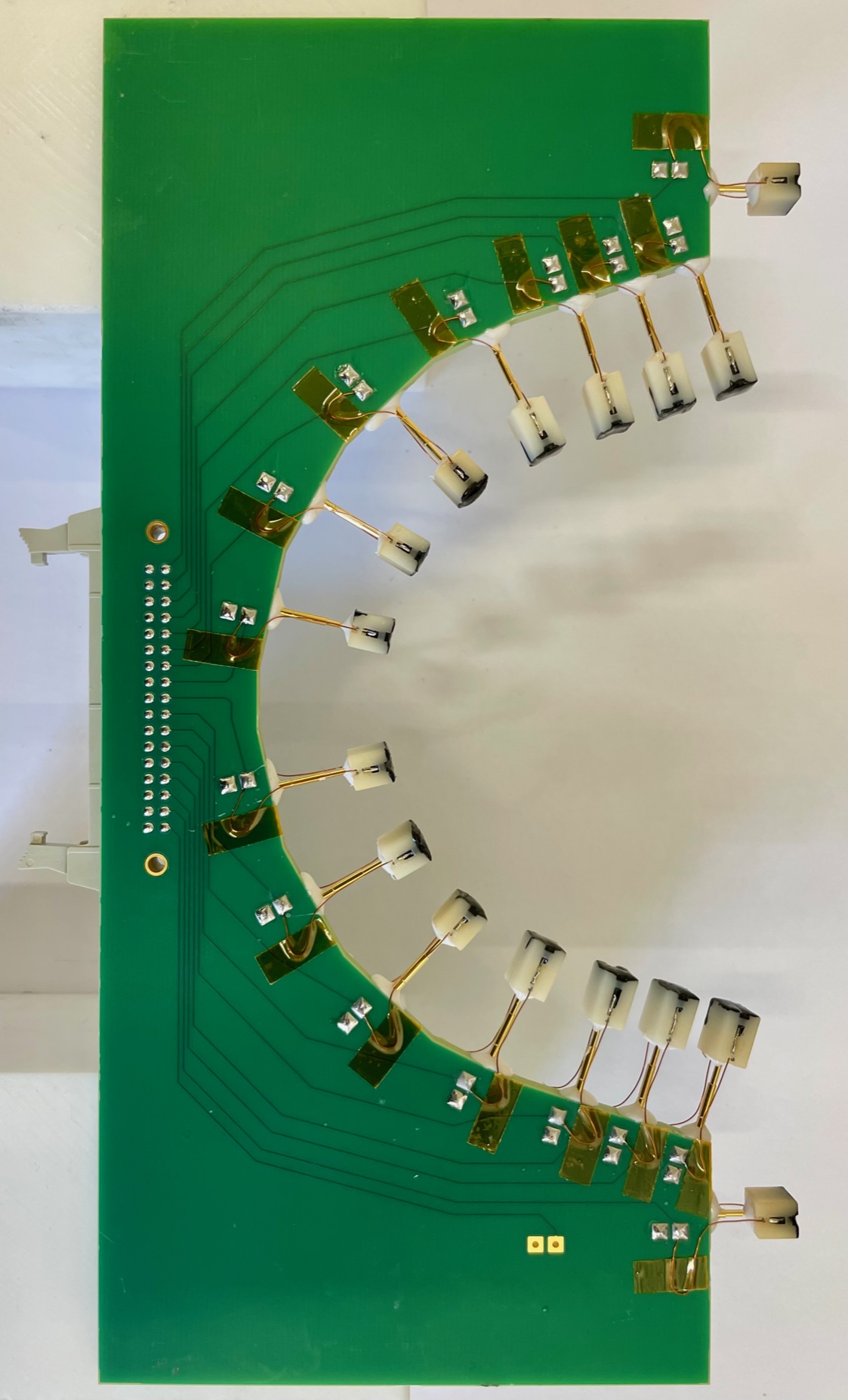} 
   \caption{One of the twelve boards used in the CERN temperature mapping system. Each board with sixteen thermometers is machined in order to fit the shape of 1.3 GHz TESLA cavities.}
   \label{fig:plot0}
\end{figure}

Thermometers are placed in contact with the cavity surface by twelve printed circuit boards (PCBs) in FR-4 glass epoxy. Each board with sixteen thermometers is adequately machined to fit the shape of a 1.3 GHz TESLA-type cavity. Each thermometer is pushed towards the cavity by a spring-loaded pin in BeCu. CAF4 silicone rubber is used to glue each spring-loaded pin to the board and the thermometer housing. This material is suitable for cryogenics and can be easily removed when a broken thermometer needs to be replaced. Concerning electrical connections, the wires from each thermometer to the electrical circuit in the PCBs are in manganin, characterized by a low thermal conductivity at liquid He temperatures. Figure \ref{fig:plot0} shows one of the boards used in the system. The density of thermometers close to the irises of the cavity cell is higher than that at the equator because the temperature profile near the irises is expected to be lower than that at the equator, as will be discussed in section \ref{sec:level5}.

\begin{figure}[!htb]
   \centering
   \includegraphics*[width=0.45\columnwidth]{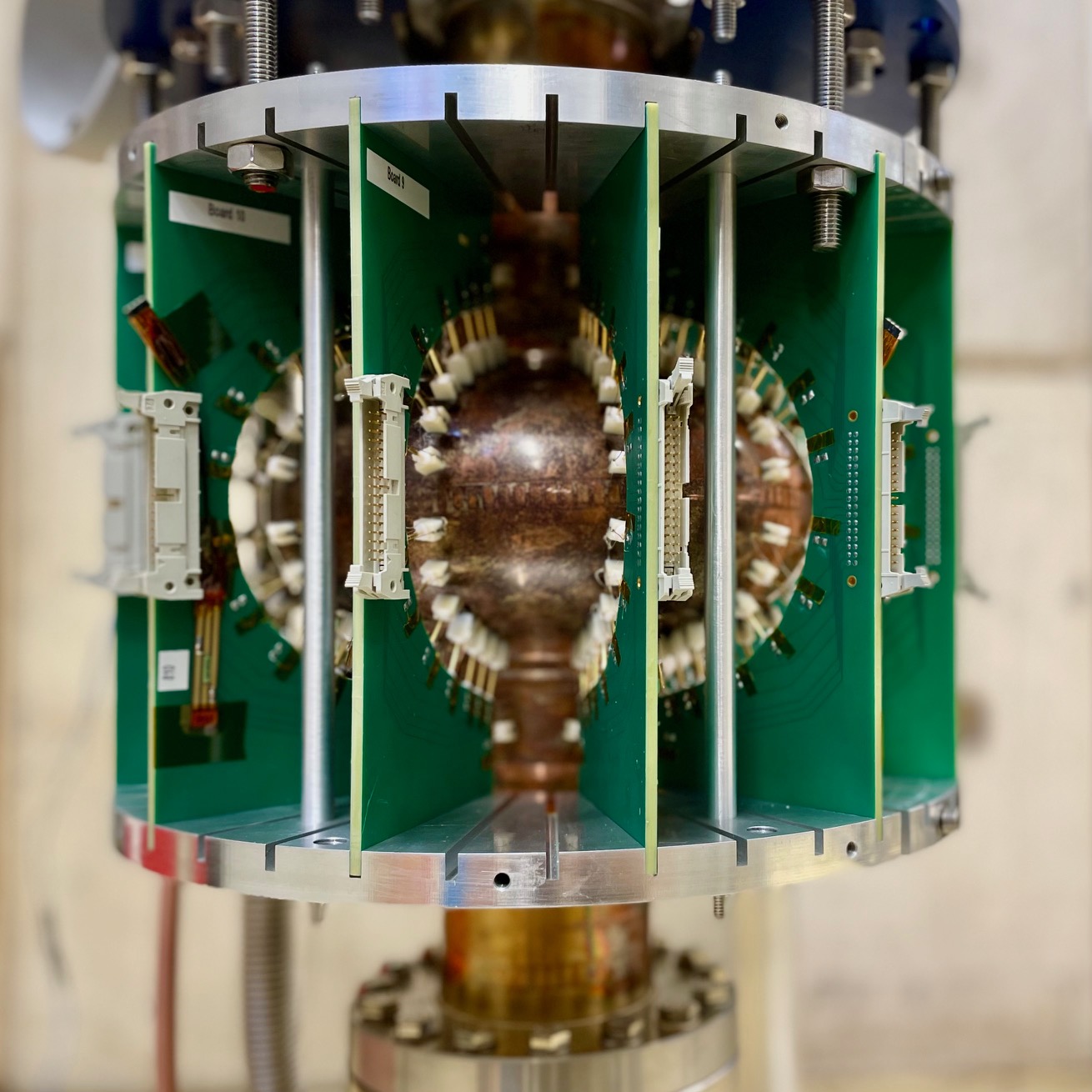} 
   \caption{Picture of the temperature mapping system assembled around a Nb/Cu 1.3 GHz single-cell cavity. The twelve thermometer boards are placed azimuthally 30 degrees apart in order to cover uniformly the whole surface of the cavity.}
   \label{fig:plot6}
\end{figure}

The supporting system, used to place the boards around the cavity under test, consists of four semicircular plates in aluminum. For each ring, 36 radial grooves are machined, where boards can be slided. For assembling the system, four locking rings are bolted on the aluminum plates after sliding the boards in the grooves and, then, the boards are pushed towards the cavity wall by set screws. The supporting system is designed to fit inside the cryostat used for tests, which has a radius of 17 cm. In the standard configuration, the twelve thermometer boards are placed azimuthally 30 degrees apart to cover the cavity uniformly, as shown in figure \ref{fig:plot6}. The spacing among the thermometers of neighboring boards varies from $\sim$5 cm close to the equator of the cavity cell and $\sim$2 cm close to the irises. Using the results of a previous study \cite{bianchi_tobepublished}, the spacing among thermometers is chosen to be enough for detecting heat losses at 2.4 K both at saturation pressure and in subcooled He.

If a limited portion of the cavity needs to be investigated with higher resolution, the boards can be placed with a minimum angle of 10 degrees to increase the density of thermometers. In this configuration, the spacing among thermometers is equal to $\sim$18 mm at the equator and $\sim$7 mm at the irises of the cavity cell.

\subsection{\label{sec:level3e2}Electronics}

\begin{figure}[!htb]
   \centering
   \includegraphics*[width=0.45\columnwidth]{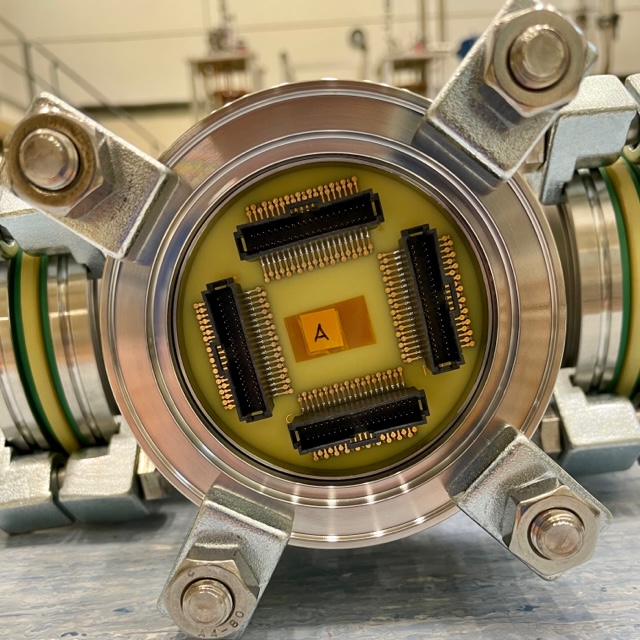} 
   \caption{Picture of one 4-mm-thick feedthrough PCB that interface the thermometers inside the cryostat and the data acquisition system on the air side.}
   \label{fig:plot7}
\end{figure}

On the top plate of the cryostat, three 4-mm-thick PCBs are used as feedthroughs to interface the thermometers inside the cryostat and the data acquisition system on the air side. The helium leak rate of these feedthrough PCBs is lower than 10$^{-8}$ mbar$\cdot$l/s. Figure \ref{fig:plot7} shows one of the feedthrough PCBs.

On both sides of each feedthrough PCB, eight 40-pin IDC connectors are soldered, as shown in figure \ref{fig:plot7}. The thermometer boards are connected to the IDC connectors on the vacuum side of feedthrough PCBs by twelve flexible CICOL 30 AWG Micro IDC cables. The additional heat load in the He bath due to those cables is evaluated as lower than 1 W. On the air side of feedthrough PCBs, the electrical signals from thermometers are transmitted to the data acquisition system by twelve standard shielded jacketed ribbon cables with 34 conductors each. Cables inside the cryostat are 3 m long, while those outside the cryostat are 7 m long. The number of conductors in ribbon cables and feedthrough PCBs is intentionally oversized to have enough redundancy in the system.

\subsection{\label{sec:level3e3}Data acquisition}

The data acquisition system consists of twelve PCBs that interface the 192 thermometers with six multiplexer switch modules (National Instruments PXIe-2527). After multiplexing, the analog signals from thermometers are digitized by six analog-digital converter modules (National Instruments NI-9251).

Each thermometer is connected to the interface PCBs with two 150 k$\Omega$ resistors in series. The thermometers are all connected in parallel, and a Keithley 2401 voltage source provides the power supply. The voltage difference is usually set to 3 V to supply a $\sim$10 $\mu$A current for each thermometer. The voltage source is connected to a computer by a general-purpose interface bus to remotely control the power supply. To avoid issues concerning the self-heating of thermometers, a current value lower than $\sim$25 $\mu$A is recommended \cite{conway2017instrumentation, canabal2007development}.

The 32-channel 2-wire multiplexer switch modules are used for multiplexing the analog signals of thermometers. Each module is capable of multiplexing 32 differential inputs. The output of each multiplexer is connected to one channel of a 24-bit analog-digital converter (ADC) module to digitize the voltage drop of each thermometer, connected in input to the multiplexer switch module. Analog signals can be digitized by these ADC channels in the range between $-$3 V and $+$3 V with high resolution. The quite large input voltage range of these ADC modules avoids significant uncertainties due to nonlinearities during the digitization of thermometer signals. Indeed, the typical voltage drop of thermometers is expected to be between $\sim$10 mV and $\sim$70 mV at liquid He temperatures. In addition, the ADC channels are equipped with mini-XLR cables to minimize noise during the data acquisition. The acquisition time of each thermometer response and the sampling rate of ADCs can be opportunely tuned to map the cavity's temperature with satisfactory accuracy. The maximum sampling rate of ADCs is 100 kS/s. Due to the high sampling rate of ADCs, high-speed temperature mapping with a maximum of twelve thermometers can also be performed by bypassing the multiplexing stage. High-speed temperature mapping allows us to study the dynamic behavior of heat losses. 

Four Temati carbon ceramic sensors, calibrated from 1.5 K to 300 K and immersed in the He bath, are connected to two ADC modules. The response of these calibrated thermometers is acquired by the 4-wire sensing method, whereas the power supply is provided by a Keithley 2401 voltage source, which is only dedicated to these four thermometers. The four carbon ceramic sensors are used to calibrate all Allen-Bradley carbon resistors at each cool-down of the system. Indeed, during the pumping on the He bath, the voltage drop of all thermometers is measured at $\sim$50 mK intervals. In order to calibrate all thermometers of the system, voltage measurements of each thermometer are correlated to the He bath temperature at the time when the thermometer response is acquired. The calibration curve of each thermometer is obtained by interpolating data with a third-order polynomial function \cite{ciovati2005temperature}.

Once the calibration is completed, the temperature measured by all thermometers is acquired without power in the cavity. Then, during the cavity test, the temperature of each thermometer is continuously acquired and subtracted from that measured when no energy was stored in the cavity. This allows the acquisition of real-time temperature maps on the cavity surface during RF tests. The acquisition time of one single temperature map can vary from 35 ms to more than 100 s, depending on the desired accuracy.

The data acquisition system is about 3 m from the cryostat outside the radiation shielding. Therefore, multiplexers and ADC modules are protected from potential radiation damage. 

As mentioned, the CERN temperature mapping system comprises Allen-Bradley 100 $\Omega$ carbon resistors. Even though a large quantity of these resistors can be found in the market, their production ceased in 1997. Therefore, the availability of Allen-Bradley carbon resistors is expected to be limited in the far future. We designed the CERN temperature mapping system for Nb/Cu cavities by considering the possibility of substituting Allen-Bradley carbon resistors with Temati carbon ceramic sensors. In such an event, the electronics and the data acquisition system are already designed to be fully compatible with the new type of thermometers.

\section{\label{sec:level4}RF PERFORMANCE OF THE CAVITY}

A Nb/Cu 1.3 GHz TESLA-type cavity was tested in a vertical cryostat at different temperatures and accelerating fields. To evaluate the RF losses of the cavity under test, the main observable is the cavity’s unloaded quality factor $Q_{0}$ as a function of accelerating field, measured by standard methods with a phase lock loop circuit \cite{padamsee2}. A mobile input coupler is used to maintain critical coupling of the cavity during the test. The measurement of $Q_{0}$ as a function of accelerating field is carried out by sweeping the input power while maintaining constant the temperature of the He bath.

$Q_{0}$ values of the cavity are measured at 2.4 K both at saturation pressure and in subcooled condition. In parallel to the evaluation of RF performance, temperature maps are acquired for several values of accelerating field ranging from $\sim$2 MV/m to $\sim$7 MV/m. The heat losses of the cavity became high enough at $\sim$2 MV/m to be detected by thermometers, whereas the measurement of $Q_{0}$ is stopped at $\sim$7 MV/m in accordance with the safety regulations of the CERN facility where the test has been performed. Indeed, during each $Q_{0}$ scan, the radiation level around the cryostat was continuously monitored, and it exceeded the allowed threshold at $\sim$7 MV/m at the onset of field emission processes in the cavity.

After filling the cryostat with liquid He at $\sim$4.2 K, the temperature of the He bath is decreased by pumping on the He bath until the cryostat pressure reaches 80 mbar, corresponding to a bath temperature of 2.4 K. Using a PID pressure controller, the pressure is always kept constant during the RF cavity test within the experimental uncertainty of 2 mbar. The experimental setup allows us to characterize the cavity at 2.4 K at saturation pressure and subcooled condition. Indeed, when the He bath is at 2.4 K, the cryostat can be quickly re-pressurized to atmospheric pressure. This leaves the liquid He at a lower temperature than that corresponding to its vapor pressure. Due to the overpressure in the cryostat, the activation of nucleation sites on surfaces immersed in the liquid He bath is suppressed and, as a consequence, the nucleate boiling regime is inhibited. Heat transfer by nucleate boiling is more effective than by convection cooling \cite{smith1969review}; therefore the cooling capability of the He bath in subcooled conditions is lower than that at saturation pressure \cite{piel1981cern}.

\begin{figure}[!htb]
   \centering
   \includegraphics*[width=0.55\columnwidth]{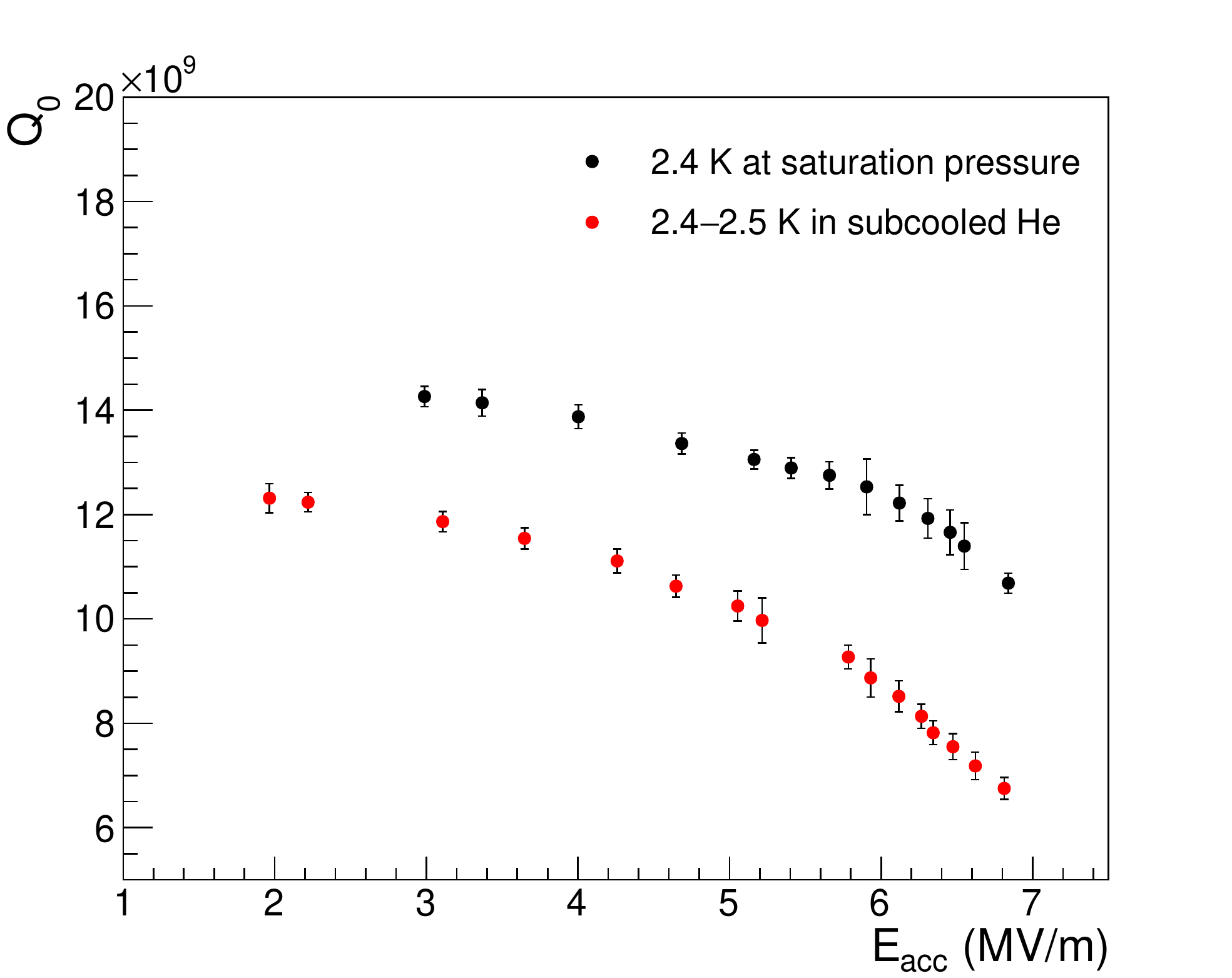} 
   \caption{Quality factor $Q_{0}$ of the cavity under test as a function of accelerating field $E_{acc}$ at the temperature of 2.4 K at saturation pressure (black) and in the temperature range between 2.4 K and 2.5 K in subcooled He (red).}
   \label{fig:plot8}
\end{figure}

Figure \ref{fig:plot8} shows the quality factor $Q_{0}$ of the cavity as a function of accelerating field $E_{acc}$ at 2.4 K at saturation pressure and in subcooled He. As demonstrated in a previous study \cite{bianchi_tobepublished}, these two He bath conditions ensure satisfactory temperature measurements along a Nb/Cu 1.3 GHz cavity cell. In both cases, progressive performance degradation with the accelerating field is observed in the cavity under test. At 2.4 K at saturation pressure, the $Q_{0}$ decreases by $\sim$20\% from 3 MV/m to 7 MV/m, while the decrease is more than 40\% at $\sim$2.4 K in subcooled He. In the subcooled condition, the temperature of the He bath is not stable over time; in fact, the temperature slowly drifts to higher values. For this reason, the $Q_{0}$ scan acquired in subcooled He is carried out in a temperature range between 2.4 K and 2.5 K. 

As shown in figure \ref{fig:plot8}, $Q_{0}$ values of the cavity in subcooled He are lower than those at 2.4 K at saturation pressure. This is mainly due to two reasons. Firstly, the $Q_{0}$ scan in subcooled He is taken when the temperature of the He bath is between 2.4 K and 2.5 K, whereas the $Q_{0}$ at saturation pressure is measured precisely at 2.4 K. The temperature increase of the He bath implies a higher surface resistance of the superconducting film and, in turn, a lower quality factor \cite{padamsee, padamsee2}. Secondly, the heat dissipation from the cavity to the He bath in the subcooled condition is generally lower than when the He bath is at saturation pressure. Because of the limited heat transfer into the He bath, the superconducting film in subcooled He turns out to be warmer than that at saturation pressure for the same accelerating field and temperature of the He bath. As a consequence, the surface resistance of the film increases and the $Q_{0}$ value decreases.

\section{\label{sec:level5}TEMPERATURE MAPS AT 2.4 K AT SATURATION PRESSURE}

\begin{figure*}[!htb]
   \centering
   \includegraphics*[width=0.95\textwidth]{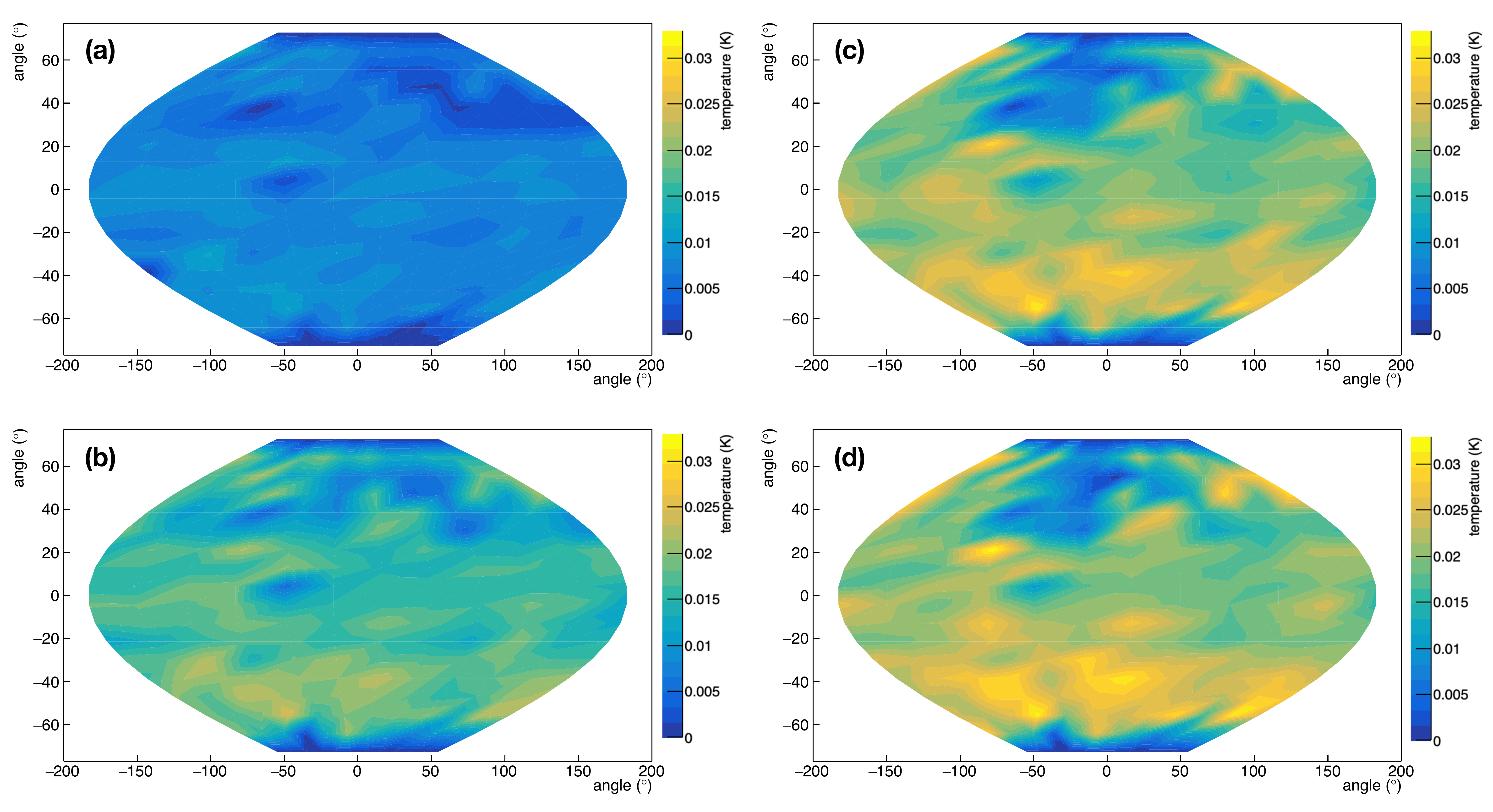}
   \caption{Temperature maps of the Nb/Cu cavity, under test, acquired at 3.0 MV/m (a), 4.6 MV/m (b), 5.8 MV/m (c), and 6.8 MV/m (d). Experimental data are taken at 2.4 K when the He bath is at saturation pressure. The abscissa of each plot represents the azimuthal coordinate around the cavity. The equator of the cavity cell is located at 0 degrees in the vertical axis of each plot, whereas the iris at the bottom of the cavity cell is at $-$75 degrees and the one at the top is at $+$75 degrees.}
   \label{fig:plot9}
\end{figure*}

For evaluating heat losses of the cavity under test, we acquired some temperature maps around the whole surface of the cavity cell at different values of accelerating fields. This section only discusses the temperature maps taken at 2.4 K when the He bath is at saturation pressure. Figures \ref{fig:plot9}a, \ref{fig:plot9}b, \ref{fig:plot9}c, and \ref{fig:plot9}d show the temperature maps acquired at 3.0 MV/m, 4.6 MV/m, 5.8 MV/m and 6.8 MV/m, respectively. The abscissa of each plot in figure \ref{fig:plot9} represents the azimuthal coordinate around the cavity. The equator of the cavity cell is located at 0 degrees in the vertical axis of each plot in figure \ref{fig:plot9}, whereas the iris at the bottom of the cavity cell is at $-$75 degrees and the one at the top is at $+$75 degrees.

As expected, the outer surface of the cavity cell is warmer by increasing the accelerating field. Indeed, the average temperature of the cavity surface is $\sim$8 mK at 3 MV/m, $\sim$14 mK at 4.6 MV/m, $\sim$16 mK at 5.8 MV/m and $\sim$20 mK at 6.8 MV/m. Some hotspots are detected in the lower half-cell, whereas a cold area is observed in the upper half-cell, as clearly shown in figures \ref{fig:plot9}c and \ref{fig:plot9}d. 

\begin{figure}[!htb]
   \centering
   \includegraphics*[width=0.55\columnwidth]{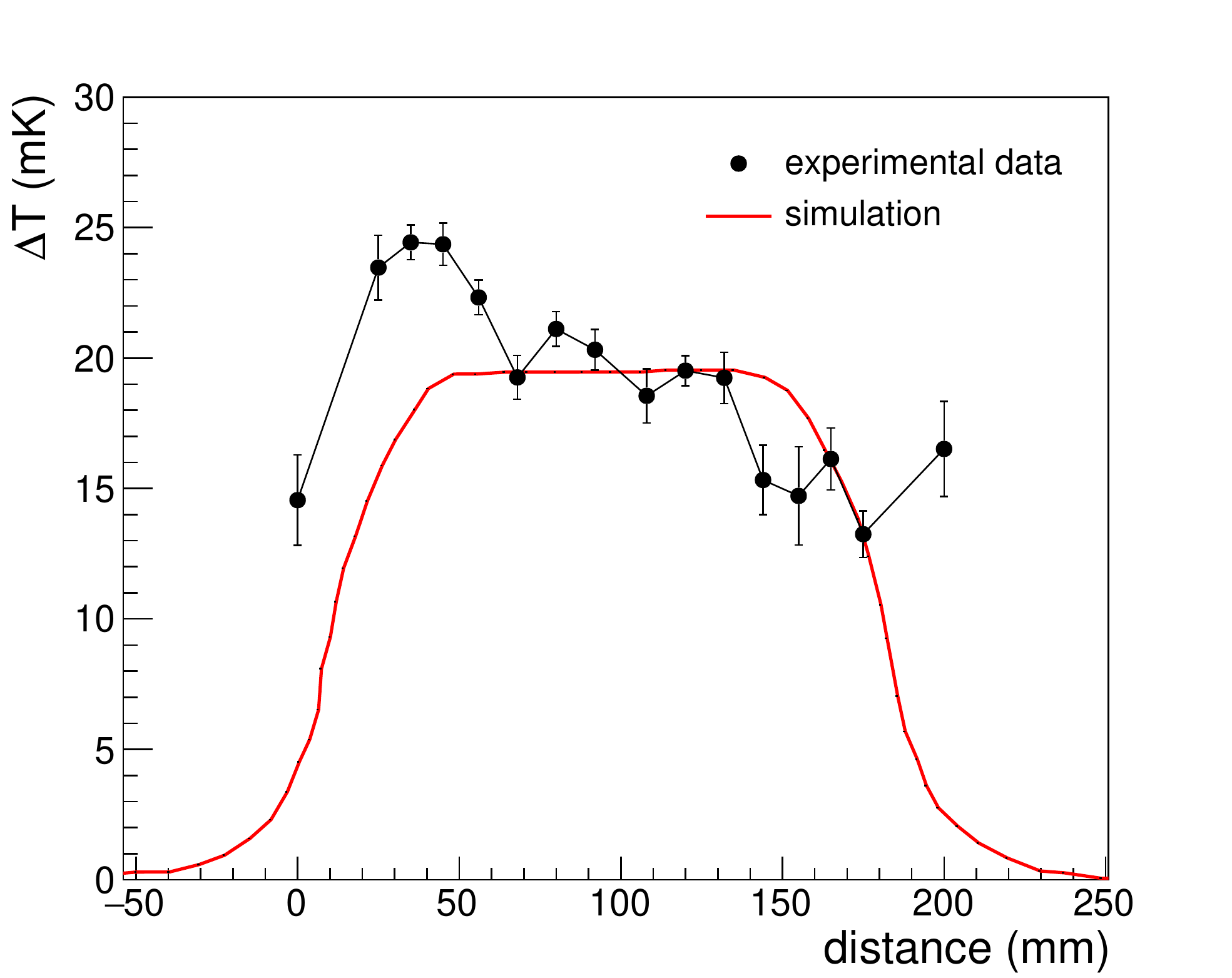} 
   \caption{Measured (black) and simulated (red) temperature profile along the meridian of the cavity cell. The equator of the cavity is at 100 mm while the lower and upper irises are at 0 and 200 mm, respectively. Experimental data are taken at 2.4 K at saturation pressure while the cavity is operated at the accelerating field of $\sim$6.4 MV/m with a quality factor of $\sim$1.2$\cdot$10$^{10}$. The measured temperature profile is averaged over all thermometers, whereas the simulated temperature profile is conveniently scaled in order to be overlapped to the measured profile for easy comparison.}
   \label{fig:plot10}
\end{figure}

For a more accurate localization of warm and cold sites, the temperature profile along the meridian of the cavity cell is shown in figure \ref{fig:plot10}. Experimental data are averaged over all thermometers of the system. The averaged temperature profile is measured at 2.4 K at saturation pressure, whereas the accelerating field of the cavity is $\sim$6.4 MV/m with a quality factor of $\sim$1.2$\cdot$10$^{10}$. The equator of the cavity is at 100 mm, while the lower and upper irises are at 0 and 200 mm, respectively.

In addition to the measured temperature profile of the cavity cell, figure \ref{fig:plot10} shows the expected temperature profile if the heat transfer from the cavity to the He bath is uniform along the whole cavity surface and the surface resistance of the Nb film does not change over the RF surface. Indeed, the power dissipated per unit area $P_{diss}$ at the RF surface in a superconducting cavity is given by:
\begin{equation}
P_{diss} = \frac{1}{2} R_{s}(T) |\textbf{H}|^{2}
\label{eq1}
\end{equation}
where $R_{s}(T)$ is the surface resistance, which depends on the temperature $T$ of the film, and $|\textbf{H}|$ is the magnitude of the RF magnetic field. Thus, if we assume that the heat transfer from the outer cavity surface to the He bath and the surface resistance of the superconducting film are uniform, the expected temperature profile is proportional to $|\textbf{H}|^{2}$ in good approximation. The profile of $|\textbf{H}|^{2}$ along the cavity cell is maximum in the equator and minimum close to the edges of the cavity cell. This profile, simulated by the software COMSOL, is conveniently scaled in figure \ref{fig:plot10} with the aim of overlapping the measured temperature profile for easy comparison.

Hotspots and cold sites can be detected by comparing the expected and measured temperature profiles in figure \ref{fig:plot10}. The lower half-cell between 20 mm and 60 mm is warmer than expected. This is explained by the presence of several hotspots, as shown in figure \ref{fig:plot9}. On the contrary, the average temperature measured between 140 mm and 160 mm is slightly lower than expected. Indeed, a cold area in the upper half-cell is observed at 2.4 K at saturation pressure, as shown in figure \ref{fig:plot9}. This area is investigated in detail in section \ref{sec:level6e1}. Finally, the average temperature measured at 0 mm and 200 mm is higher than expected. Thermometers at 0 mm and 200 mm measure the temperature in correspondence with two welds at the edges of the cavity cell. These two welds could locally modify the substrate's thermal properties and induce imperfections on the Cu surface where the Nb thin film is deposited. As a consequence, the heat dissipation of the Cu substrate and the superconducting behavior of the thin film could be altered in the correspondence of both welds. This might explain why the welds are warmer than expected.

\subsection{\label{sec:level5e2}Characterization of heat losses}

\begin{figure}[!htb]
   \centering
   \includegraphics*[width=0.65\columnwidth]{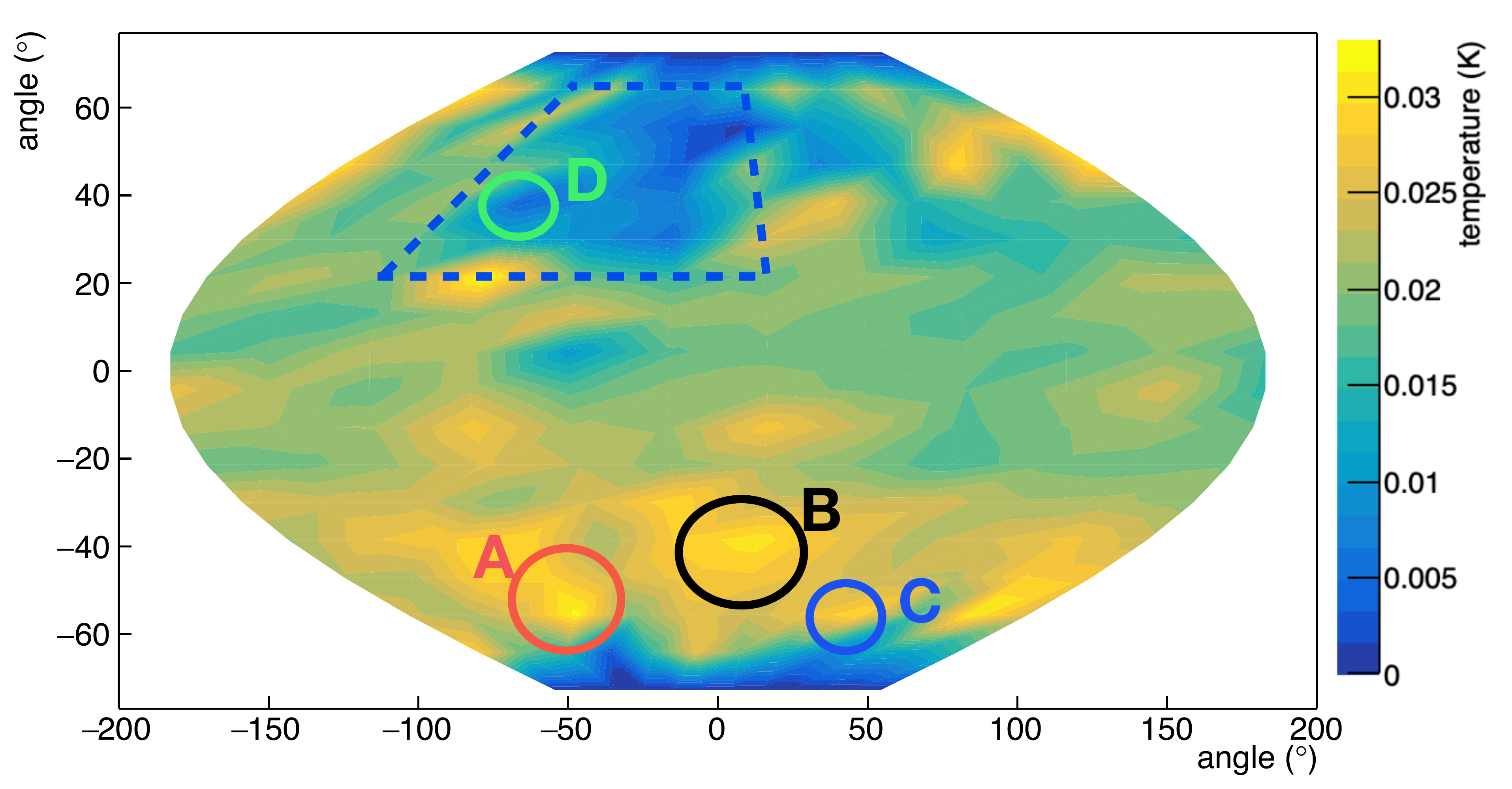} 
   \caption{Temperature map of the Nb/Cu cavity under test, acquired at $\sim$6.8 MV/m with a $Q_{0}$ value of $\sim$1.1$\cdot$10$^{10}$. Four different sites of interest are labeled by A, B, C and D, while a cold area in the upper half-cell is enclosed by the blue dashed curve. The abscissa of the plot represents the azimuthal coordinate around the cavity. The equator of the cavity cell is located at 0 degrees in the vertical axis of the plot, whereas the iris at the bottom of the cavity cell is at $-$75 degrees and the one at the top is at $+$75 degrees.}
   \label{fig:plot11}
\end{figure}

We can characterize the type of heat losses observed in the temperature maps. In particular, we can identify some sites of the cavity that show behavior consistent with Joule heating or field emission heating. Figure \ref{fig:plot11} shows the temperature map of the cavity at 2.4 K at saturation pressure. The temperature map is acquired at $\sim$6.8 MV/m with a $Q_{0}$ value of $\sim$1.1$\cdot$10$^{10}$.

In figure \ref{fig:plot11}, four different areas of interest are labeled as A, B, C, and D, while a blue dashed line encloses the cold region. Hotspots A, B, and C are located in the lower half-cell and represent some of the warmest sites of the cavity during the test. An additional site, labeled by D, is located inside the cold area and shows an interesting behavior, as will be discussed in the following.

\begin{figure}[!htb]
   \centering
   \includegraphics*[width=0.55\columnwidth]{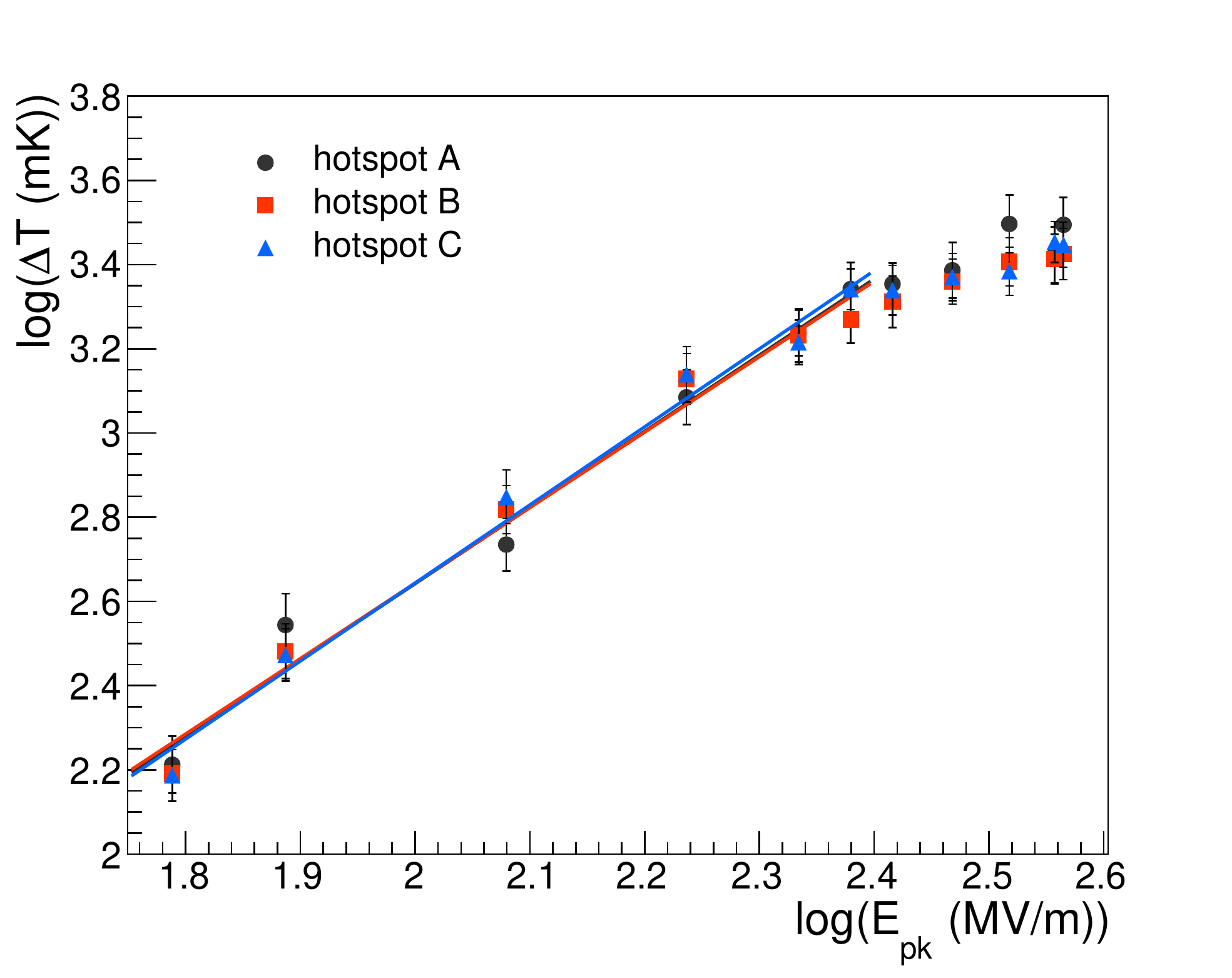} 
   \caption{Temperature variation $\Delta T$ as a function of the peak electric field $E_{pk}$ in correspondence of hotspots A, B, and C, according to the notation of figure \ref{fig:plot11}. Experimental data in the log-log plot are taken at 2.4 K at saturation pressure. Straight lines interpolate data for values of $E_{pk}$ lower than $\sim$10 MV/m. Markers hide some statistical error bars.}
   \label{fig:plot12}
\end{figure}

If we assume that the heat transfer from the cavity to the He bath and the surface resistance of the superconducting film are uniform, the temperature rise measured close to a heat loss with ohmic behavior scales as the square of peak electric field $E_{pk}^{2}$ \cite{padamsee2}. Figure \ref{fig:plot12} shows the temperature variation $\Delta T$ as a function of $E_{pk}$ in a log-log plot in correspondence of hotspots A, B, and C, according to the notation of figure \ref{fig:plot11}.

By interpolating the experimental data, we found that the slope of lines in figure \ref{fig:plot12} is 1.81 $\pm$ 0.12 for hotspot A, 1.79 $\pm$ 0.12 for hotspot B and 1.85 $\pm$ 0.11 for the hotspot C when $E_{pk}$ is lower than $\sim$10 MV/m. If the measured temperature rise is due purely to Joule heating, the slope is equal to 2 \cite{padamsee2}. The statistical z-test confirms that all three lines that interpolate the data in figure \ref{fig:plot12} have a slope compatible with the expected value of 2. Therefore, the behavior of hotspots A, B, and C is ohmic for $E_{pk}$ values lower than $\sim$10 MV/m. For values of $E_{pk}$ higher than $\sim$10 MV/m, the temperature rise in all three hotspots does not depend on the square of the peak electric field anymore. This can be correlated to the transition of the cold area from convection cooling to nucleate boiling, as discussed in section \ref{sec:level6e1}.

\begin{figure}[!htb]
   \centering
   \includegraphics*[width=0.55\columnwidth]{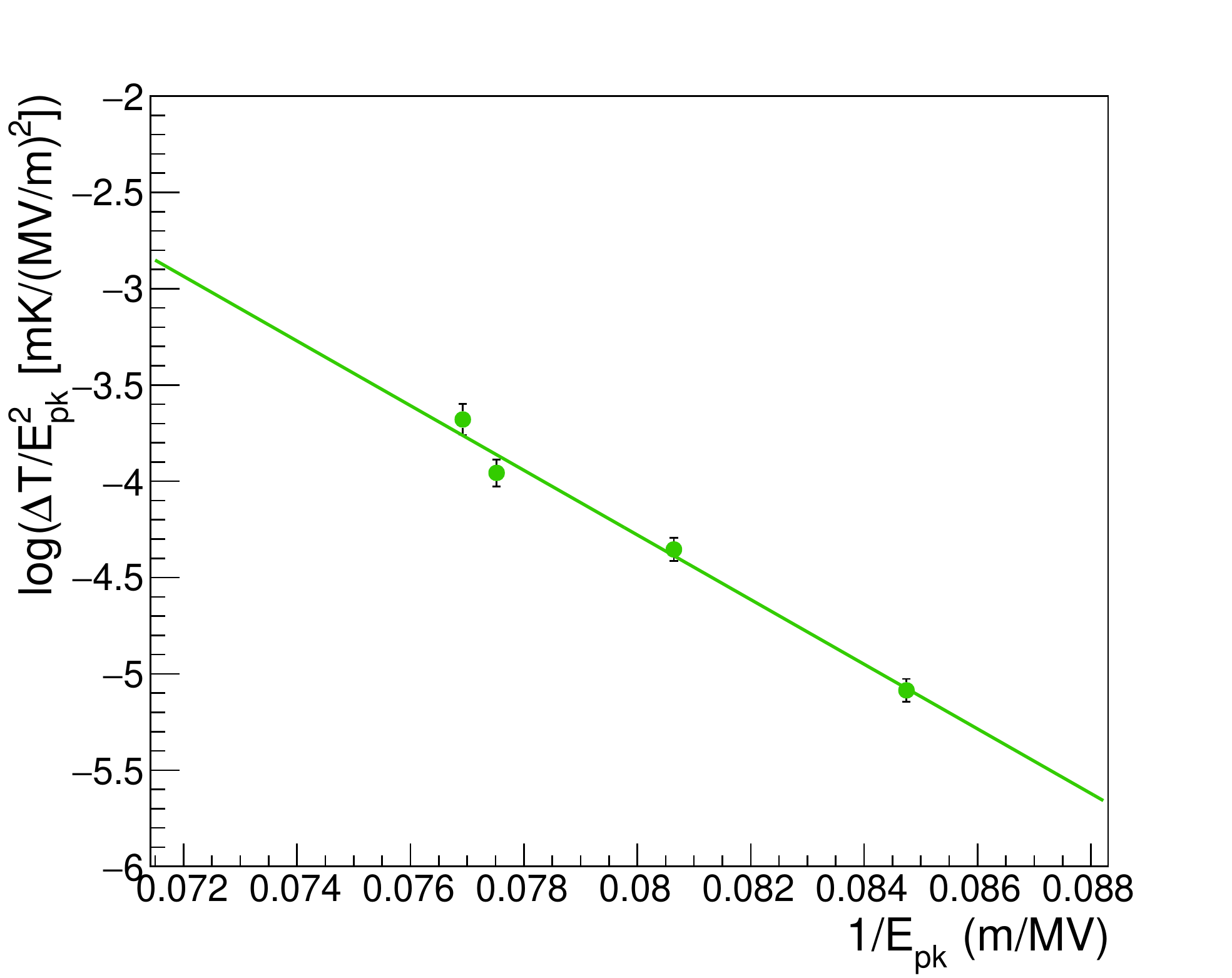} 
   \caption{Fowler-Nordheim plot in correspondence of site D, according to the notation of figure \ref{fig:plot11}. In Fowler-Nordheim plots, ln($\Delta T$/$E_{pk}^{2}$) is plotted as a function of 1/$E_{pk}$, where $\Delta T$ is the temperature variation and $E_{pk}$ is the peak electric field. Experimental data are acquired at 2.4 K at saturation pressure. Data interpolation is done by a straight line for values of $E_{pk}$ between $\sim$12 MV/m and $\sim$13 MV/m.}
   \label{fig:plot13}
\end{figure}

Temperature measurements of the cavity surface can localize field emitters using Fowler-Nordheim plots, where ln($\Delta T$/$E_{pk}^{2}$) is plotted as a function of 1/$E_{pk}$. Figure \ref{fig:plot13} shows the Fowler-Nordheim plot in correspondence with site D at 2.4 K at saturation pressure. The linear behavior in the Fowler-Nordheim plot indicates that a field emitter is present in correspondence with site D in the cavity under test. The slope of the line that interpolates data in figure \ref{fig:plot13} can be used to determine the effective emission area. A complete description of field emission in SRF cavities and Fowler-Nordheim plots is beyond the scope of this study. The reader is referred to the literature for further details \cite{padamsee, padamsee2}.

\section{\label{sec:level6}TEMPERATURE MAPS AT $\sim$2.4 K IN SUBCOOLED HELIUM}

\begin{figure*}[!htb]
   \centering
   \includegraphics*[width=0.95\textwidth]{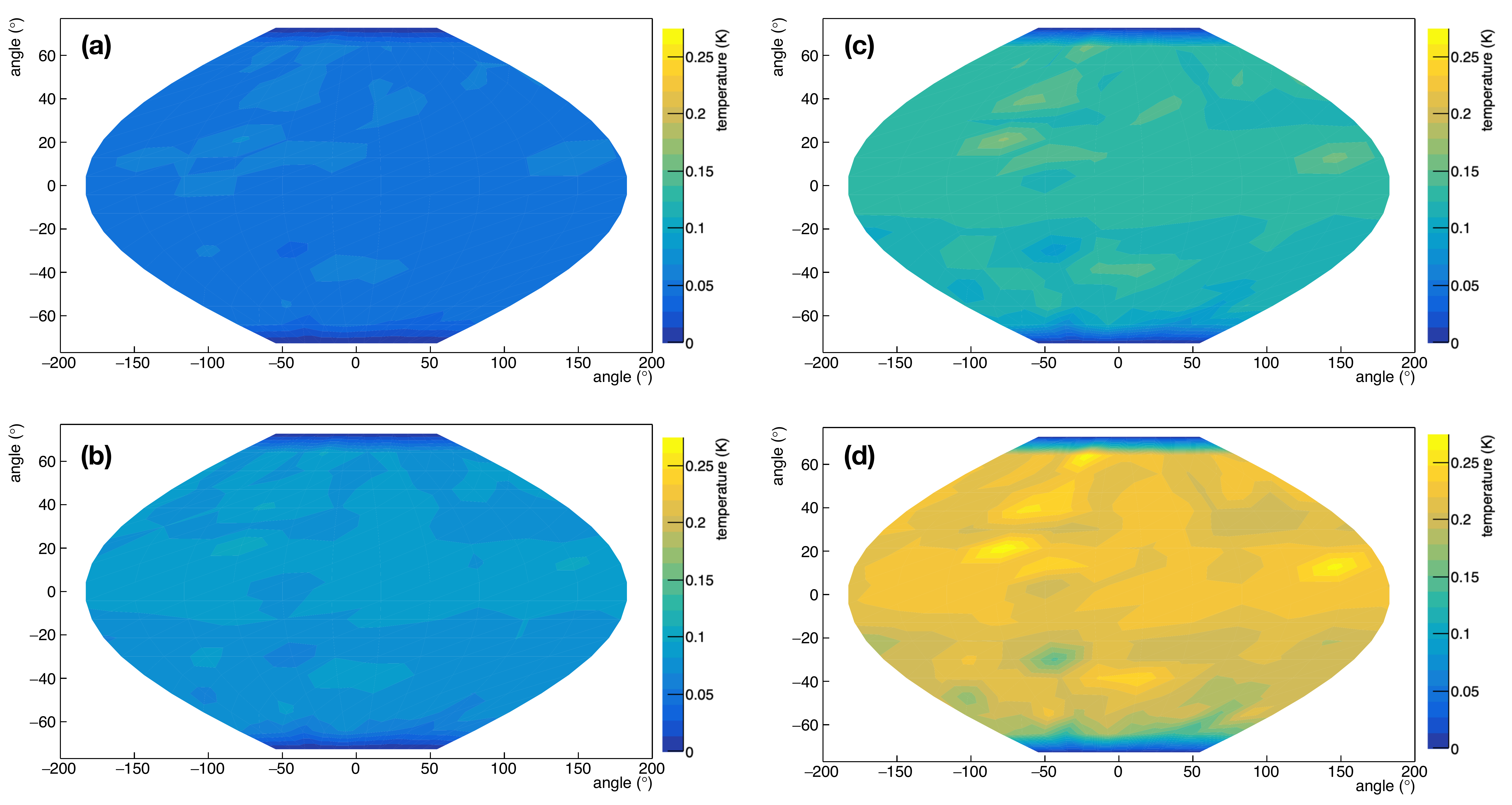}
   \caption{Temperature maps of the Nb/Cu cavity, under test, acquired at 3.0 MV/m (a), 4.6 MV/m (b), 5.8 MV/m (c), and 6.8 MV/m (d). Experimental data are taken at $\sim$2.4 K when the He bath is in the subcooled condition. The abscissa of each plot represents the azimuthal coordinate around the cavity. The equator of the cavity cell is located at 0 degrees in the vertical axis of each plot, whereas the iris at the bottom of the cavity cell is at $-$75 degrees and the one at the top is at $+$75 degrees.}
   \label{fig:plot14}
\end{figure*}

Nb/Cu cavities are usually operated in a liquid He bath at saturation pressure. However, for a better characterization of the cold area, we also acquired temperature maps at $\sim$2.4 K in subcooled He. The heat transfer from the cavity to the He bath in subcooled He is less effective than that at saturation pressure because the cooling mainly occurs by convection. Indeed, the overpressure over the liquid He bath impedes the onset of the nucleate boiling regime. Figures \ref{fig:plot14}a, \ref{fig:plot14}b, \ref{fig:plot14}c and \ref{fig:plot14}d show the temperature maps at $\sim$2.4 K in the subcooled condition acquired at 3.0 MV/m, 4.6 MV/m, 5.8 MV/m and 6.8 MV/m, respectively. Unlike temperature maps at 2.4 K at saturation pressure, no cold area is observed at the same He bath temperature in the subcooled condition. 

\subsection{\label{sec:level6e1}Nucleate boiling regime}

\begin{figure}[!htb]
   \centering
   \includegraphics*[width=0.5\columnwidth]{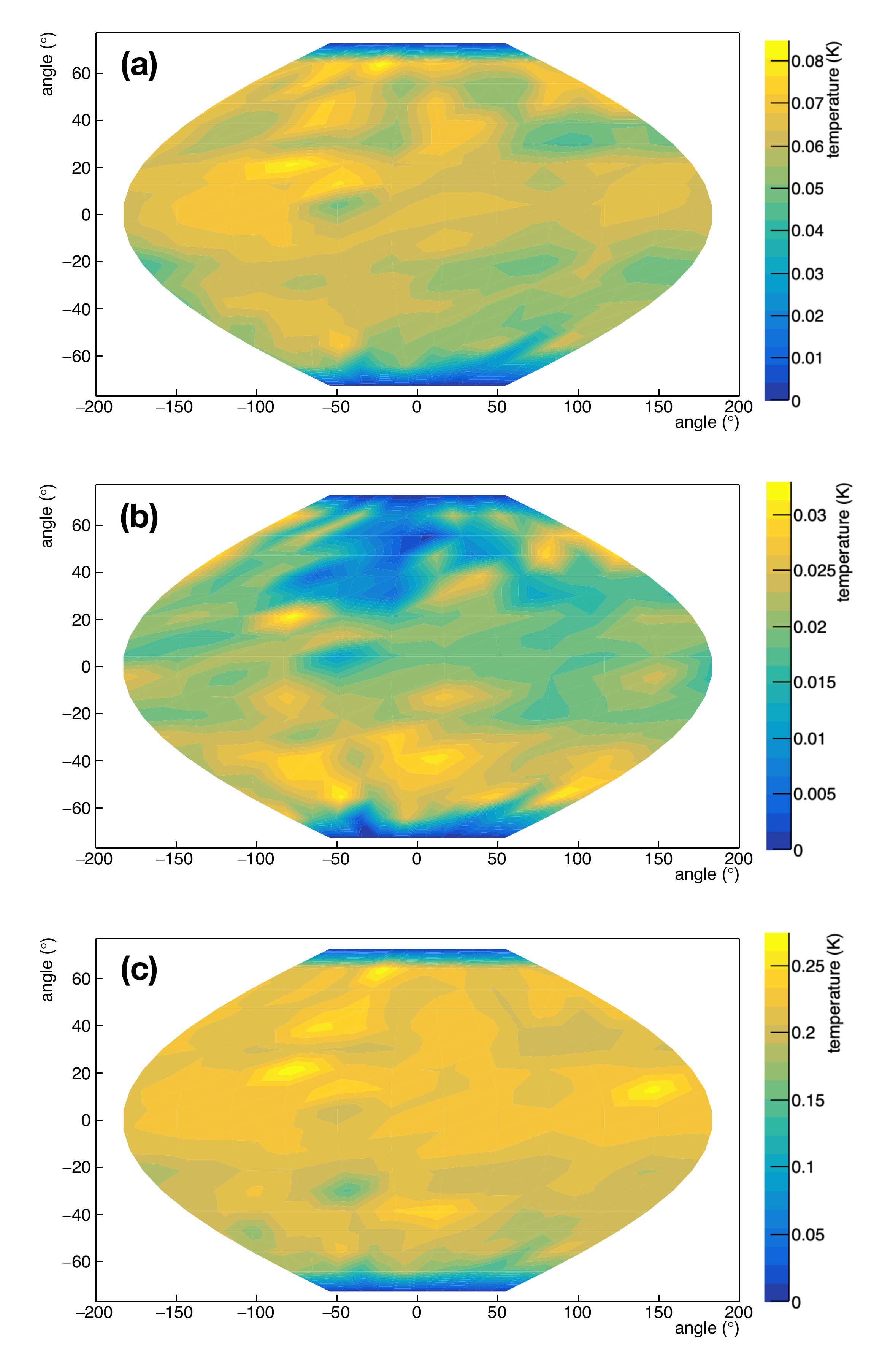} 
   \caption{Temperature maps of the cavity under test, sequentially acquired at $\sim$6.8 MV/m when the He bath is at $\sim$2.2 K in the subcooled condition (a), then at 2.4 K at saturation pressure (b) and, finally, at $\sim$2.4 K in the subcooled condition (c).}
   \label{fig:plot15}
\end{figure}

To better understand the origin of the cold area, a systematic acquisition of temperature maps at different temperatures of the He bath, both at saturation pressure and in subcooled He, was carried out to dismiss the hypothesis of any potential issues during data collection. Temperature maps were sequentially acquired by changing only the He bath conditions. Figures \ref{fig:plot15}a, \ref{fig:plot15}b and \ref{fig:plot15}c show the temperature maps acquired at $\sim$6.8 MV/m when the He bath is at $\sim$2.2 K in the subcooled condition (a), then at 2.4 K at saturation pressure (b) and, finally, at $\sim$2.4 K in the subcooled condition (c). The temperature of the He bath in subcooled conditions has been intentionally chosen at temperatures lower and higher than that at saturation pressure. We observed that the cold area appears only at saturation pressure and, consequently, is not due to a potential issue of the experimental setup we used for data acquisition. 

Since the cold area is observed only at saturation pressure, as shown in figure \ref{fig:plot15}, the heat might be dissipated in the cold area by nucleate boiling. In other words, a portion of the cavity could be colder than the remaining cavity surface because it is cooled by nucleate boiling, which is more effective than just convection cooling. However, this cannot occur when the He bath is in the subcooled condition because the nucleate boiling regime is impeded by the overpressure over the He bath. We also observed that site D, according to the notation of figure \ref{fig:plot11}, is the warmest site in the cold area when the He bath is in the subcooled condition.

\begin{figure}[!htb]
   \centering
   \includegraphics*[width=0.55\columnwidth]{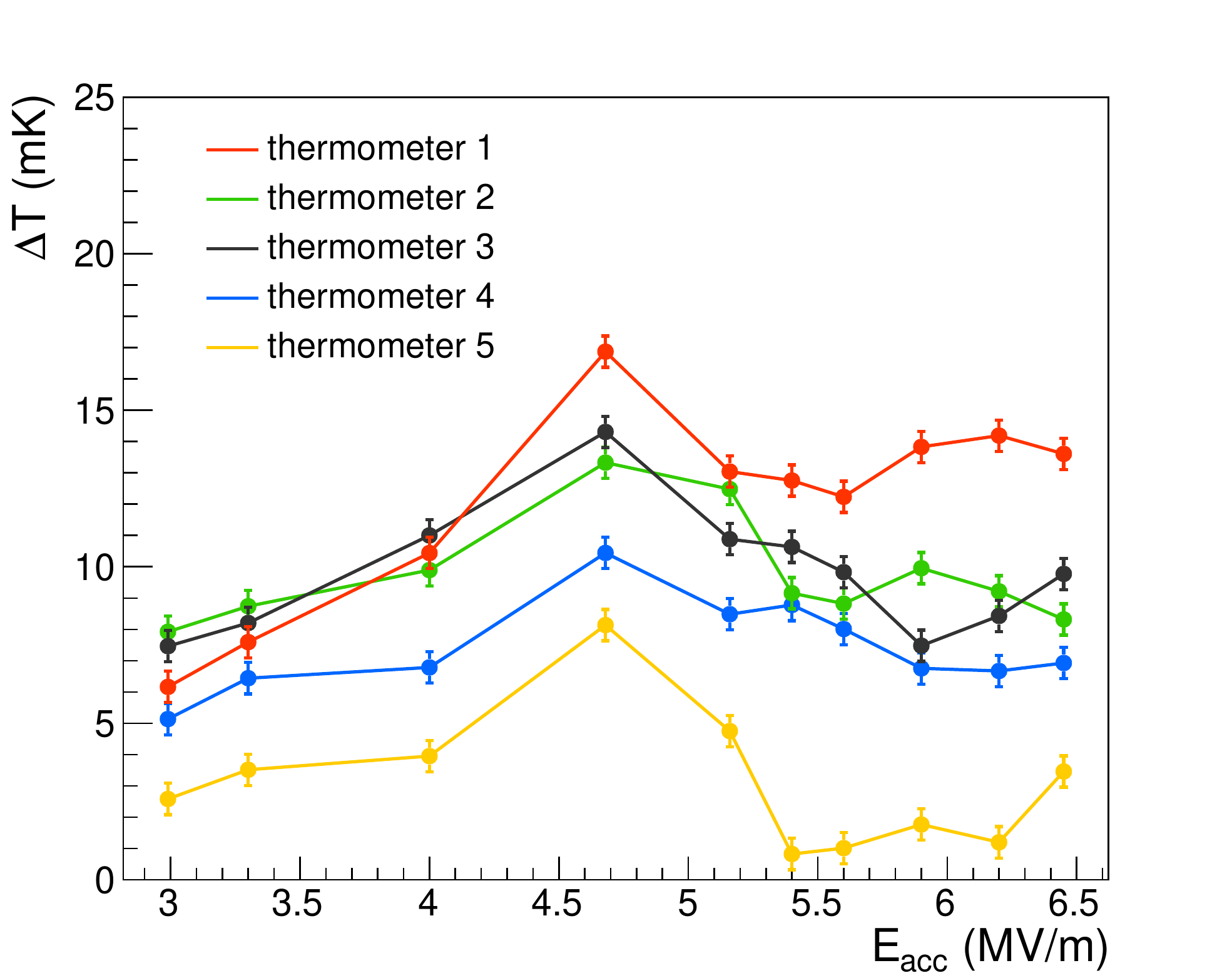} 
   \caption{Temperature variation $\Delta T$, measured by five thermometers located in correspondence of the cold area, as a function of accelerating field $E_{acc}$ at 2.4 K when the He bath is at saturation pressure.}
   \label{fig:plot16}
\end{figure}

In addition to the observation that the cold area is only present at saturation pressure, a temperature drop is measured by most of the thermometers in correspondence with this area for a particular value of power in the cavity. Figure \ref{fig:plot16} shows the temperature variation, measured by five thermometers located in correspondence of the cold area, as a function of accelerating field $E_{acc}$ at 2.4 K when the He bath is at saturation pressure. A global temperature drop is detected for accelerating field values  higher than $\sim$5 MV/m. This provides further evidence to support our hypothesis. Indeed, the transition from convection cooling to nucleate boiling is associated with a temperature drop due to the activation of nucleation sites \cite{smith1969review}.

Our findings suggest that the cavity cell is cooled by convection except for one portion where the transition from convection cooling to nucleate boiling might have occurred. 

\section{\label{sec:level7}OPTICAL INSPECTION OF THE CAVITY}

\begin{figure}[!htb]
   \centering
   \includegraphics*[width=0.50\columnwidth]{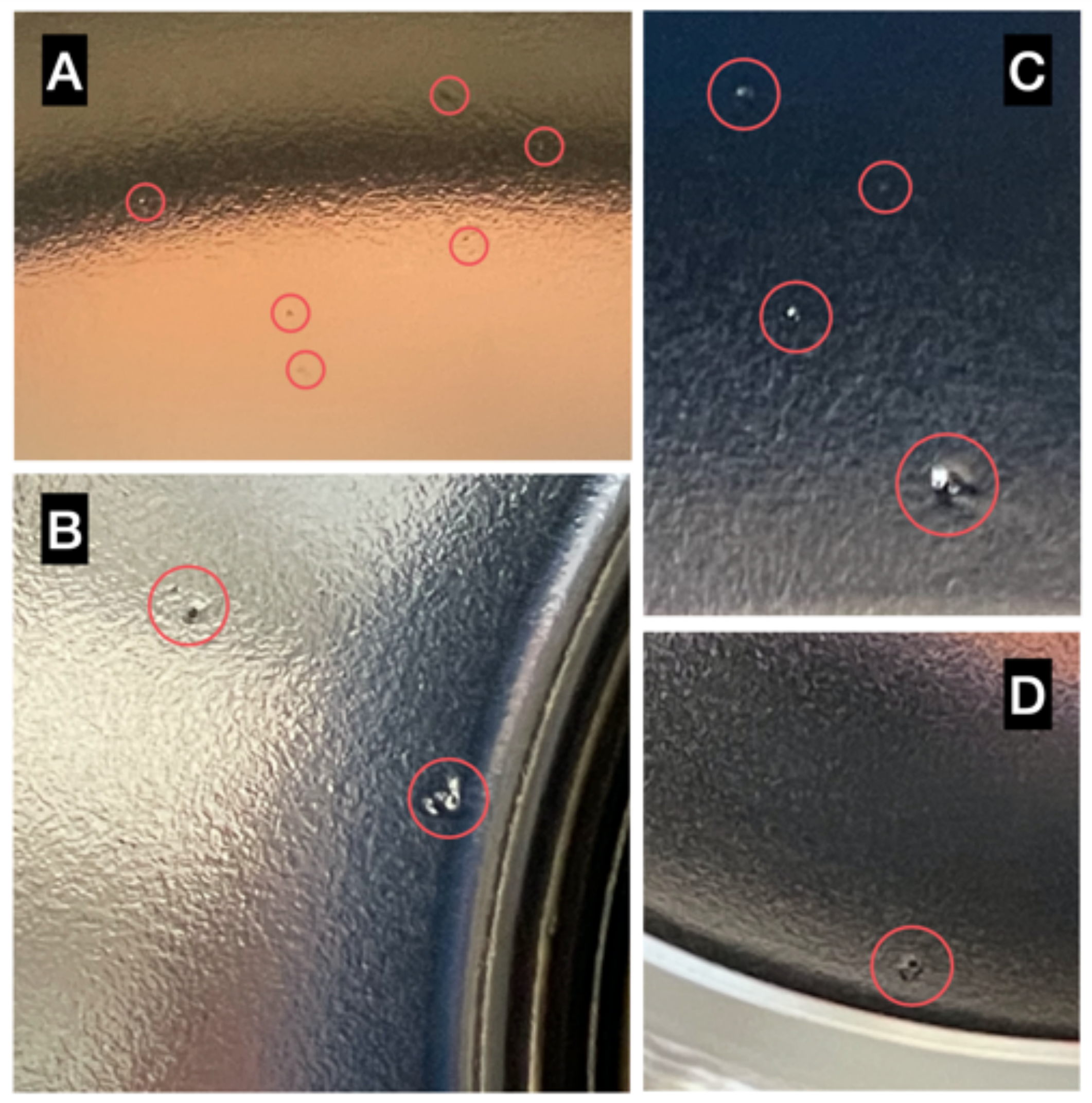} 
   \caption{Defects of the Nb thin film in correspondence of sites A, B, C, and D, according to the notation of figure \ref{fig:plot11}.}
   \label{fig:plot17}
\end{figure}

Through an optical inspection of the inner cavity surface, we observed several defects of the Nb film in correspondence with hotspots detected in the temperature maps. We carried out the optical inspection using a commercial camera. Only a limited portion of the inner surface was accessible during the inspection; in fact, we intentionally avoided inserting the camera inside the cavity to avoid damaging it and, if necessary, retesting it for further investigations. Figure \ref{fig:plot17}a, \ref{fig:plot17}b, and \ref{fig:plot17}c show the defects in the Nb film in correspondence with hotspots A, B, and C, respectively. These sites display characteristics consistent with Joule heating, as demonstrated in section \ref{sec:level5e2}. Figure \ref{fig:plot17}d shows an additional defect observed exactly in correspondence to site D. This site shows a behavior consistent with field emission heating, therefore the defect shown in figure \ref{fig:plot17}d could be the field emitter of the cavity. Indeed, during the cold tests, RF measurements of the cavity were stopped at $\sim$7 MV/m because of the high radiation level in the test facility due to field emission. No significant differences are visually observed between the defects with ohmic behavior in figures \ref{fig:plot17}a, \ref{fig:plot17}b and \ref{fig:plot17}c, and the defect with behavior consistent with field emission heating in figure \ref{fig:plot17}d.

\begin{figure}[!htb]
   \centering
   \includegraphics*[width=0.5\columnwidth]{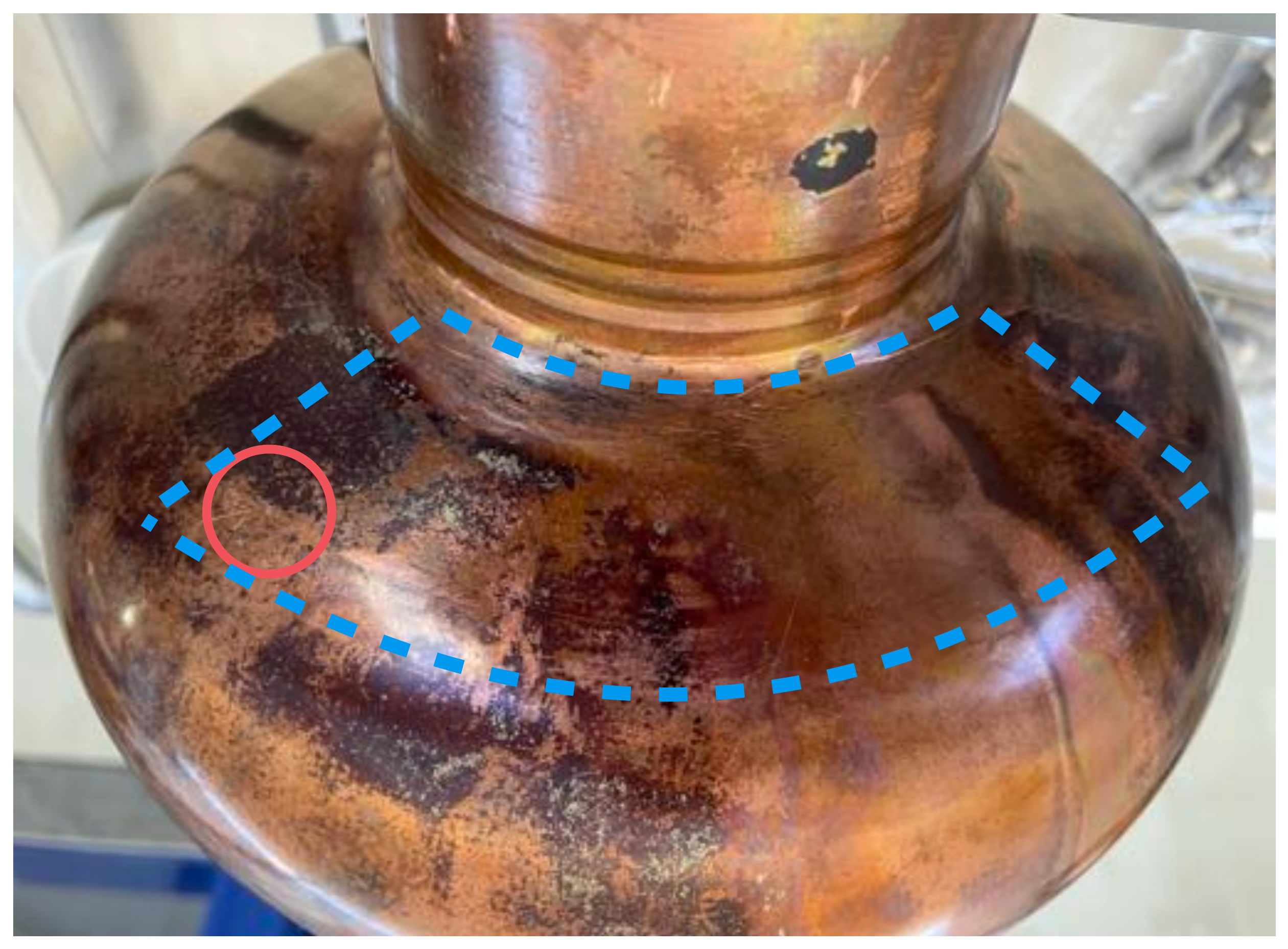} 
   \caption{Picture of the outer surface of the Nb/Cu cavity in correspondence with the cold area, which is observed during the cavity tests at 2.4 K at saturation pressure.}
   \label{fig:plot18}
\end{figure}

Figure \ref{fig:plot18} shows the cavity's outer surface in correspondence with the cold area. The area that appears cold at 2.4 K at saturation pressure during the cavity test is surrounded by a dashed curve in blue. No significant differences between this area and the remaining surface of the cavity cell are observed by optical inspection.

\section{\label{sec:level8}HEAT DISSIPATION OF NIOBIUM-COATED COPPER CAVITIES}

Using our temperature mapping system, we observed that the heat dissipation of the Nb/Cu cavity under test is not uniform over the whole surface at operating conditions. Temperature maps in figure \ref{fig:plot9} clearly show that a cold area is present on the outer surface of the Nb/Cu cavity when operated in a He bath at saturation pressure. As discussed in section \ref{sec:level6e1}, several hints suggest that this cold area is in nucleate boiling regime. To dissipate the heat of the cavity surface into the He bath, cooling by nucleate boiling turns out to be more effective than convection. This finding might suggest a way to improve RF performance for Nb/Cu cavities.

Under normal operating conditions, the surface resistance of the superconducting Nb film increases exponentially with temperature \cite{halbritter1970comparison}. Therefore, limiting the temperature rise of the superconducting Nb films during the operation of SRF cavities implies higher RF performance. For a constant heat flux, enhanced heat transfer into the He bath implies a lower temperature of the RF surface in Nb/Cu cavities that are usually operated in He-I between 4.0 K and 4.5 K at saturation pressure, where the thermal contribution in the surface resistance is relatively high.

A number of authors observed that the heat transfer from Cu surfaces to liquid He-I is generally affected by several factors \cite{smith1969review, bald1976nucleate, schmidt1981review}. One of those parameters is the roughness of the Cu surface. According to data in the literature \cite{smith1969review, bald1976nucleate, schmidt1981review, bianchi_tobepublished}, the heat transfer into He-I bath is higher in rough Cu surfaces, where more heat can be dissipated, than that in smooth surfaces.

If the cooling by nucleate boiling regime can be induced in large areas of Nb/Cu cavities by increasing their external roughness, the heat exchange from Nb/Cu cavities to the He-I bath may be enhanced and, in turn, their RF performance improved. Increasing the roughness of the outer surface of Nb/Cu cavities can be easily achieved by several manufacturing processes. For example, sandblasting could be a straightforward and cheap process that can provide uniform and reproducible results in terms of roughness in large Cu surfaces.

\section{\label{sec:level9}CONCLUSIONS}

In this paper, we have described a temperature mapping system based on contact thermometry and designed to effectively sense heat losses in Nb/Cu 1.3 GHz single-cell TESLA-type cavities in He-I both at saturation pressure and in subcooled conditions. Temperature maps in He-I at saturation pressure are acquired to study the heat dissipation from the cavity to the He bath at operating conditions, whereas temperature maps in subcooled conditions allow the localization of hotspots with high precision and accuracy.

Temperature mapping of a 1.3 GHz Nb-coated cavity indicates that the heat dissipation is not uniform over the whole external cavity surface. Indeed, heat losses are localized and can have behavior consistent with Joule heating or field emission heating. In addition, our findings suggest that the cavity under test is mainly cooled by convection cooling, except for one region that is in the nucleate boiling regime.

\section*{\label{sec:level13}ACKNOWLEDGEMENTS}
We gratefully acknowledge the contribution of our colleagues Giovanna Vandoni, Pablo Vidal Garcia, and Lorena Vega Cid for their support of the project. We warmly thank Marco Chiodini, Florence Crochon and Agostino Vacca for their invaluable help.

\section*{\label{sec:level14}DATA AVAILABILITY}
The data that support the findings of this study are available from the corresponding author upon reasonable request.

\nocite{*}

\bibliography{apssamp}

\end{document}